\documentclass[12pt,preprint]{aastex}

\newcommand{\pfrac}[2]{\frac{\partial #1}{\partial #2}}
\newcommand{\kakkoi}[3]{\left(\frac{{#1}}{{#2}}\right)^{#3}}

\shorttitle{Final mass of giant planets}
\shortauthors{Tanigawa \& Ikoma}

\begin{document}
%\baselineskip=10mm


%\title{Evolution and Final Mass of Gaseous Planets in Proto-Planetary
%Disks}
\title{A SYSTEMATIC STUDY OF THE FINAL MASSES OF GAS GIANT PLANETS}
%\title{Systematic study of final mass of gas giant planets}


%\author{Takayuki Tanigawa and Masahiro Ikoma\altaffilmark{1}}
%\author{Takayuki Tanigawa\altaffilmark{1} and Masahiro Ikoma}
%\affil{Earth and Planetary Sciences, Tokyo Institute of Technology,
%    Tokyo, Japan}

\author{Takayuki Tanigawa and Masahiro Ikoma}
\affil{Department of Earth and Planetary Sciences, 
Tokyo Institute of Technology, Ookayama, Meguro-ku, Tokyo 152-8551, 
Japan}
\email{tanigawa@geo.titech.ac.jp}

%\and

%\author{Masahiro Ikoma \altaffilmark{}}
%\affil{Research Center for the Evolving Earth and Planets, 
%Tokyo Institute of Technology, Ookayama, Meguro-ku, Tokyo 152-8551, 
%Japan}
%\email{mikoma@geo.titech.ac.jp}


%\altaffiltext{1}{tanigawa@geo.titech.ac.jp}


\begin{abstract}
We construct an analytic model for the rate of gas accretion onto a
planet embedded in a protoplanetary disk as a function of planetary
mass, disk viscosity, disk scale height, and unperturbed surface density
in order to study the long-term accretion and final masses of gas giant
planets.
We first derive an analytical formula for surface density profile near
the planetary orbit from considerations of the balance of force and the
dynamical stability.  Using it in the empirical formula linking surface
density with gas accretion rate that is derived based on hydrodynamic
simulations of Tanigawa \& Watanabe (2002, ApJ 586, 506), we then
simulate the mass evolution of gas giant planets in viscously-evolving
disks.
We finally determine the final mass as a function of semi-major axis of
the planet.  We find that the disk can be divided into three regions
characterized by different processes by which the final mass is
determined.
In the inner region, the planet grows quickly and forms a deep gap to
suppress the growth by itself before disk dissipation.  The final mass
of the planet in this region is found to increase with the semi-major
axis in a similar way to the mass given by the viscous condition for gap
opening, but the former is larger by a factor of approximately 10 than
the latter.
In the intermediate region, viscous diffusion of the disk gas limits the
gas accretion before the planet form a deep gap.  The final mass can be
up to the disk mass, when disk viscous evolution occurs faster
than disk evaporation.
In the outer region, planets capture only tiny amounts of gas within the
lifetime of the disk to form Neptune-like planets.
We derive analytic formulae for the final masses in the different
regions and the locations of the boundaries, which are helpful to gain
a systematic understanding of the masses of gas giant planets.

\end{abstract}


\keywords{methods: analytical --- methods: numerical --- solar system: formation}



%%%%%%%%%%%%%%%%%%%%%%%%%%%%%%%%%%%%%%%%%%%%%%%%%%%%%%%%%%%%%%%%%%%%%%%
\section{Introduction}
A fundamental but unresolved issue with planet formation is how the mass
of a giant planet is fixed.  In the solar system there are four
giant planets, which are characterized by their massive hydrogen/helium
envelopes. Jupiter and Saturn consist mostly of hydrogen/helium, while
Uranus and Neptune are mostly of ice but have significant amounts of
hydrogen/helium.  The giant planets are different in mass: Jupiter's
mass is $1 \times 10^{-3} M_\odot (= M_J)$, Saturn's mass is $\sim 0.3
M_J$, and Uranus' and Neptune's masses are $\sim 0.05 M_J$.
Furthermore, the extrasolar planets detected so far also range in mass
from about one Neptune mass to $\sim$
10~$M_J$\footnote{http://exoplanet.eu/}; those planets are believed to
have massive hydrogen/helium envelopes like the giant planets in the
solar system.

% Core accretion model vs. Disk instability model
There are two competing ideas for the formation of giant planets, the
core accretion model \citep[e.g.,][]{Miz80,BP86} and the disk
instability model \citep[e.g.,][]{Cameron78,Boss89}.  In the core
accretion model, a rocky/icy core first forms through collisional
aggregation of planetesimals, followed by envelope formation due to
substantial accretion of gas from the circumstellar (protoplanetary)
disk.  In the disk instability model, density fluctuation of the disk
gas grows to form a gaseous planet, followed by core formation due to
sedimentation of heavy elements in its interior. The advantages and
disadvantages of both models are discussed in several literatures
\citep[e.g.,][]{Boss02}, which is not repeated here.  In this paper we
consider giant planet formation in the context of the core accretion
model.

% Subcritical or supercritical phases
In the core accretion model, the process of the accumulation of the
envelopes is divided into two phases, the \textit{subcritical-accretion}
and \textit{supercritical-accretion} phases, in terms of the dominant
energy source.  The transition from the former to the latter occurs when
the mass of a core reaches a critical value.  In the
subcritical-accretion phase, incoming planetesimals supply energy to the
envelope so that the envelope is in the hydrostatic equilibrium. The core
accretion thus controls the gas accretion in this phase.  The phase 2
found by \citet{Pol96} is a part of the subcritical-accretion phase. In
the supercritical-accretion phase, the energy supplied by planetesimals
is insufficient to keep the hydrostatic structure of the envelope, so
that the envelope substantially contracts and releases its gravitational
energy, resulting in runaway accretion of the disk gas.

% envelope-limited or disk-limited subphases in the supercritical phase
The supercritical-accretion phase can be further divided into two
subphases.  In the former subphase, the gas accretion is controlled by
contraction of the envelope and thus occurs on the Kelvin-Helmholtz
timescale \citep{BP86,Pol96,INE00,IG06}.  However, because the gas
accretion driven by the envelope contraction accelerates rapidly with
time, the supply of the disk gas inevitably becomes unable to keep up
with the demand of the contracting envelope. Thus, in the latter
subphase, the disk-gas supply limits the gas accretion
\citep[][hereafter TW02]{TW02}.

% The phase we focus on
In this paper we focus on the gas accretion in the latter subphase of
the supercritical-accretion phase, i.e., the phase in which the
Kelvin-Helmholtz contraction of the envelope is under way but the gas
accretion is limited by disk-gas supply. After the onset of the
runaway gas accretion (i.e., in the supercritical-accretion phase), the
planet always experiences the limited gas supply (i.e., the latter
subphase) by the end of its growth. Although there is an idea that the
masses of Uranus and Neptune were fixed in the subcritical-accretion
phase because of dissipation of the disk gas \citep{Pol96}, we will
explore another possibility that the masses of those planets were
fixed in the supercritical-accretion phase.

% Gap formation
One idea is that the mass of a giant planet is completely fixed when the
planet opens a gap in the disk. From a physical point of view, the gap
opens when the gravitational scattering by the planet overwhelms both
viscous diffusion (the viscous condition) and push due to pressure
gradient (the thermal condition) \citep[e.g.,][]{LP93}. In normal
protoplanetary disks, the final mass of a giant planet is determined by
the thermal condition \citep[e.g.,][]{IL04}, which is equivalent to the
condition that the Hill radius is equal to the disk scale height at the
planet's location. The final mass is thus determined only by temperature
of the disk gas for given stellar mass and planet's location.  In
protoplanetary disks similar to the minimum-mass solar nebula
\citep{Hay81}, the thermal condition yields the final mass of
approximately one Jupiter mass at 5~AU, but predicts that the masses of
giant planets increase with the semi-major axis, which is in clear
contradiction to the current configuration of the solar
system. Furthermore, because, in reality, the gap is not vacuum but a
low-density region, continuous gas accretion through the gap takes
place, as suggested by several studies \citep[e.g.,][]{AL96}. The
effects of the subsequent accretion on the final masses of giant planets
have been poorly understood.

% Depleted disk
Another idea is that gas accretion is truncated by depletion of the disk
gas before gap opening (e.g., viscous dissipation and photo-evaporation
of the disk gas). This means that the final mass of a giant planet is
the total mass of the disk gas that the planet has captured until the
disk gas dissipates completely. This may be responsible for the current
configuration of the solar system, since it takes longer time for outer
giant planets to accrete principally because of their long orbital
periods.  However, there have been no quantitative studies of this
possibility.

% Purpose
The purpose of this paper is to gain a systematic understanding of how
the final mass of a giant planet is determined. Specifically, we clarify
(1) how much the mass of an accreting giant planet increases via slow,
continuous accretion through the gap, (2) when and where disk
dissipation dominates gap opening, and finally (3) how the final masses
of giant planets depend on several disk parameters. To do so, we
simulate the long-term accretion of a giant planet in a self-consistent
fashion. The essential ingredients of the modeling are global viscous
evolution of the disk and the flow pattern of the accreting gas, as
described below; especially, the latter effect was completely neglected
by previous studies.  Furthermore, we derive analytical formulas for the
final masses of giant planets and the boundaries between several
different regimes.

% Global evolution
During the viscous evolution of the disk, gas radially migrates from
outer regions to the site of giant planet formation. The mass flux
limits gas accretion onto giant planets in some situations. Because of
limited CPU speed of computers, azimuthally-averaged one-dimensional
simulations of the disk evolution are inevitable to follow the long-term
accretion.  Several previous studies simulated the radial mass flux in
viscously evolving disks and then calculated gas accretion rate onto
giant planets.
\citet{LS03} and \citet{Guillot06} also simply assumed that gas
accretion rate onto giant planets is always proportional to the viscous
mass flux.  \citet{VA04} and \citet{AMB04} used an approximate formula
for gas accretion rate that fits the results of two-dimensional
hydrodynamic simulations of disk gas with an embedded planet done by
\citet{LSA99} and \citet{DHK02}.  However, the formula was derived based
on a limited number of numerical simulations. Hence, it is in question
whether the formula applies in a variety of situations.

% Detailed hydrodynamic simulations
To understand the detailed pattern of accretion flow and to obtain the
gas accretion rate onto the planet, several workers
\citep[e.g.,][]{Kley99, LSA99, DHK02, Bate03, DKH03} performed two- or
three-dimensional hydrodynamic simulations of interaction between a
planet and disk gas.  Since they simulated global disks, the effect of
gap formation on accretion rate was automatically included.  However,
because of the time-consuming simulations, they did not simulate the
growth of giant planets on timescales of the viscous disk evolution.

% Tanigawa & Watanabe (2002)
This study is clearly different from the previous studies in that this
study includes the effects of the flow pattern of the accreting gas in
simulating long-term growth of planets embedded in globally evolving
disks.
To do so, we use the empirical relation between gas accretion rate and
unperturbed surface density that was derived by TW02 (see \S 3
for the details).  They carried out detailed two-dimensional
hydrodynamic simulations of gas accretion flow onto a planet.  They
demonstrated the importance of the local flow pattern on the accretion
rate.

% Structure of this paper
In this paper, we first derive an analytical formula for radial surface
density distribution of protoplanetary disks with an embedded planet in
\S 2. Next, we describe how to include the effect of the flow pattern of
accreting gas, following TW02, to calculate gas accretion rate onto
giant planets in \S 3.  In \S 4, we simulate the long-term accretion of
giant planets, based on the prescription given in \S\S 2 and 3, and
obtain the final masses of giant planets for wide ranges of several disk
parameters. Also, we derive analytical approximate formulas for the
final masses of giant planets and the boundaries between several
regimes.  Finally, we conclude this paper in \S 5.

%%%%%%%%%%%%%%%%%%%%%%%%%%%%%%%%%%%%%%%%%%%%%%%%%%%%%%%%%%%%%%%%%%%%%%%
\section{Surface density}\label{sec_surface_density}

% Description of situation we consider.
We consider a planet embedded in a gas disk, both of which are assumed
to rotate with the Keplerian velocity around a star.
We derive an approximate formula for azimuthally-averaged surface
density as a function of radial distance to the central star, $r$.  We
adopt a local coordinate system in which all quantities except the
surface density are independent of $r$.  We assume steady states and no
mass flow toward the central star.  We do not consider migration of the
planet.  The validity of these assumptions are discussed in \S
\ref{sec_remarks}.

We employ a local coordinate system co-rotating with the planet on a
circular Keplerian orbit around the central star (a local shearing-sheet
approximation).  The origin of the coordinate system is the planet's
position.  The $x$ axis is on the line from the star through the planet;
the direction of the $y$ axis is the same as that of the velocity vector
of the planet.  The equation of motion of the disk gas in the $y$
direction in this coordinate system is given by
\begin{equation}
\pfrac{}{t} (\Sigma v_y)
 +\pfrac{}{x} (\Sigma v_y v_x)
= \pfrac{}{x} \left(\nu \Sigma \pfrac{v_y}{x}\right)
 +\Sigma \dot{v}_y,
\label{EOM-y}
\end{equation}
where $\Sigma$ is surface density, $v_x$ and $v_y$ are velocities in the
$x$ and $y$ directions, $\nu$ is viscosity coefficient, and $\dot{v}_y$
is force acting on gas per unit mass.  Here we have omitted the
advection term in the $y$ direction because of axisymmetry.  We also
eliminate all the terms on the left-hand side of this equation on the
assumptions of steady states $(\partial / \partial t = 0)$ and no mass
flow $(v_x=0)$.  Hereafter we consider only $x>0$ because of symmetry.

To derive an analytical formula for $\Sigma(x)$, we assume that $v_y$ is
equal to the Keplerian velocity, $v_y = -(3/2)\Omega_{\rm p}x$ where
$\Omega_{\rm p}$ is the Keplerian angular velocity at the planet's
position.  This assumption will be found to be reasonable later.
Equation (\ref{EOM-y}) is thus transformed into
\begin{equation}
-\frac{3}{2} \nu \Omega_{\rm p}
 \pfrac{\ln \Sigma}{x}
+ \dot{v}_y
= 0.
 \label{vy_balance}
\end{equation}

The gravity of the planet perturbs the motion of the disk gas.  We
follow the impulse prescription \citep[e.g.,][]{LP79} that approximates
scattering of the disk gas as small-angle scattering of a particle by a
point-mass object.
On the prescription, $\dot{v}_y$ is given by
\begin{equation}
\dot{v}_y
 = \frac{4}{9\pi}
   \kakkoi{M_{\rm p}}{M_\ast}{2}
   r_{\rm p}^5
   \Omega_{\rm p}^2
   x^{-4},
 \label{vydot}
\end{equation}
where $r_{\rm p}$ is the distance of the planet from the central star
and $M_{\rm p}$ and $M_\ast$ are the masses of the planet and the
central star, respectively.
This approximation is inappropriate for $x \lesssim {\rm min} (2r_{\rm
H}, r_{\rm p}\sqrt{(M_{\rm p}/M_\ast)/h})$, where $r_{\rm H} \equiv
r_{\rm p}(M_{\rm p}/3M_\ast)^{1/3}$ namely, the Hill radius, because
this region is what is called the horse-shoe region, in which gas is not
pushed away from the planet.  However, since the gas in the horse-shoe
region has no contribution to the accretion flow of interest (see in \S
\ref{sec_accretion_rate}), we neglect such a region.
Note that \citet{CMM06} proposed a more elaborate formula for the torque
(a $y$-direction force in our formulation) that includes the radial
transfer of torque through waves, which they found in their
two-dimensional hydrodynamic simulations.  However, since the formula is
unable to apply for wide ranges of the parameters, as they mentioned, we
do not use their formula in this study.

Substituting equation (\ref{vydot}) into equation (\ref{vy_balance}), we
obtain
\begin{equation}
\Sigma (x)
 = \Sigma_\infty
   \exp
   \left[
    -\kakkoi{x}{\ell}{-3}
   \right]
 \equiv \Sigma_{\rm vis}(x),
\label{Sigma_vis}
\end{equation}
where $\Sigma_\infty$ is the surface density at infinity (i.e., far away
from the planet's orbital radius) and $\ell$ is defined as
\begin{eqnarray}
\ell
 &\equiv&
  \left[
   \frac{8}{81\pi}
   \kakkoi{\nu}{r_{\rm p}^2\Omega_{\rm p}}{-1}
   \kakkoi{M_{\rm p}}{M_\ast}{2}
  \right]^{1/3}
  r_{\rm p} \label{ell} \\
 &=& 0.146
  \kakkoi{\nu}{10^{-5}r_{\rm p}^2\Omega_{\rm p}}{-1/3}
  \kakkoi{M_{\rm p}}{10^{-3}M_\ast}{2/3}
  r_{\rm p}.
\label{ell_typical}
\end{eqnarray}
This surface density profile is basically the same as that derived by
\citet{LD06}, except for the value of the coefficient.

The dynamical stability of the surface density profile given by equation
(\ref{Sigma_vis}) must be checked using the well-known Rayleigh
criterion \citep[e.g.,][]{Chandra61}.  For the surface density profile
to be stable, in the local coordinate system,
\begin{equation}
\pfrac{v_y}{x} \geq -2\Omega_{\rm p}
 \label{Rayleigh_on_local}
\end{equation}
must be fulfilled.
The velocity $v_y$ can be obtained by solving the equation of motion for
the $x$ component,
\begin{equation}
   3\Omega_{\rm p}^2 x
 + 2\Omega_{\rm p} v_y
 - c^2 \pfrac{\ln \Sigma}{x}
 = 0,
\label{x_balance}
\end{equation}
where $c$ is sound speed.  Substituting equation~(\ref{Sigma_vis}) into
equation~(\ref{x_balance}), we obtain
\begin{equation}
v_y
 =-\frac{3}{2} \Omega_{\rm p} x
   \left[ 1-\kakkoi{h}{\ell}{2} \kakkoi{x}{\ell}{-5}
   \right],
\label{v_y_vis}
\end{equation}
where $h$ is the disk scale height defined as $h\equiv c/\Omega_{\rm
p}$.  From equations~(\ref{Rayleigh_on_local}) and (\ref{v_y_vis}), we
find that condition~(\ref{Rayleigh_on_local}) is not fulfilled for $x <
x_{\rm m}$, where
\begin{eqnarray}
x_{\rm m}
 &\equiv& 12^{1/5} \kakkoi{h}{\ell}{2/5} \ell
  =       12^{1/5} \kakkoi{\ell}{h}{3/5} h  \label{x_m} \\
 &=& 0.207
  \kakkoi{h}{0.1r_{\rm p}}{2/5}
  \kakkoi{\nu}{10^{-5}r_{\rm p}^2\Omega_{\rm p}}{-1/5}
  \kakkoi{M_{\rm p}}{10^{-3}M_\ast}{2/5}
  r_{\rm p}.
 \label{x_m_typical}
\end{eqnarray}
In the region $x \leq x_{\rm m}$, the density gradient thus has to be
small enough to satisfy the Rayleigh condition: the density profile
would be relaxed to be marginally stable for the Rayleigh condition.
Integrating $\partial v_y / \partial x = -2\Omega_{\rm p}$ inward from
$x_{\rm m}$ we obtain
\begin{equation}
 v_y = v_{y,\rm m} - 2\Omega_{\rm p} (x-x_{\rm m})
\end{equation}
in the region $x\leq x_{\rm m}$.  Substituting this into equation
(\ref{x_balance}), we finally obtain
\begin{equation}
\Sigma (x)
 = \Sigma_\infty
   \exp
   \left(
    -\frac{1}{2}
     \left(
      \frac{x}{h} - \frac{5}{4} \frac{x_{\rm m}}{h}
     \right)^2
    +\frac{1}{32} \kakkoi{x_{\rm m}}{h}{2}
    -\kakkoi{x_{\rm m}}{\ell}{-3}
   \right)
 \equiv \Sigma_{\rm R},
\label{Sigma_R}
\end{equation}
for $x \leq x_{\rm m}$.

\begin{figure}
\epsscale{.80}
\plotone{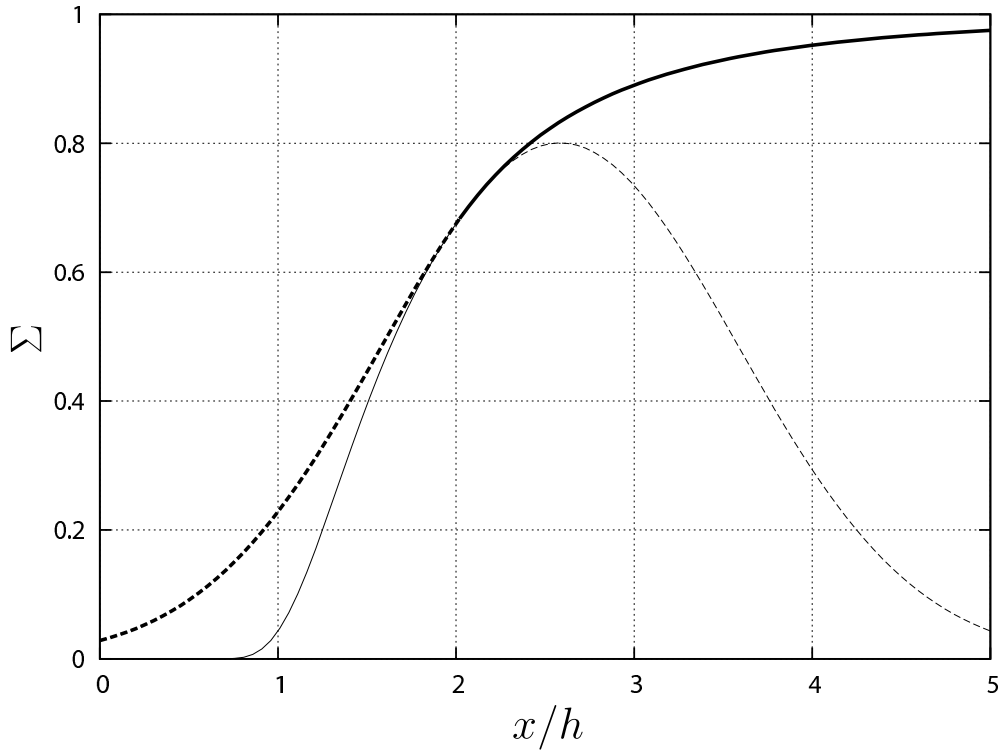}
\caption{Surface density profile for $M_{\rm p}/M_\ast = 10^{-3}$,
$\nu/(r_{\rm p}^2\Omega_{\rm p})=10^{-5}$, and $h/r_{\rm p}=0.1$.  Solid
line is $\Sigma_{\rm vis}$ given by equation (\ref{Sigma_vis}) and
dashed line is $\Sigma_{\rm R}$ given by equation (\ref{Sigma_R}).  The
two curves connect smoothly with each other at $x=x_{\rm m}$ (i.e.,
$x=2.08h$ in this case).  Thick line is the actual surface density we
use in this paper.
\label{fig_typical_sigma}}
\end{figure}

In summary, the equilibrium profile of surface density is given as
\begin{equation}
\Sigma(x)
 = \left\{
\begin{array}{ll}
 \Sigma_{\rm R}(x)   & \mbox{for } x \leq x_{\rm m}, \\
 \Sigma_{\rm vis}(x) & \mbox{for } x \geq x_{\rm m}.
\end{array}
\right.
\label{Sigma_general}
\end{equation}
An example of the density profile is shown in
Fig.\ref{fig_typical_sigma}.

As mentioned above, we have assumed $v_y = -(3/2)\Omega_{\rm p}x$ in
deriving equation (\ref{Sigma_vis}).  However, based on equation
(\ref{v_y_vis}), the velocity at $x=x_{\rm m}$ is
\begin{equation}
v_{y,\rm m}
 = -\frac{11}{8} \Omega_{\rm p} x_{\rm m}.
\end{equation}
This means that the maximum deviation from the Keplerian shear velocity,
$-(3/2)\Omega_{\rm p} x$, is less than 10\% of the absolute value.  Hence
the assumption of the Keplerian shear velocity is reasonable for $x \geq
x_{\rm m}$.

\begin{figure}
\epsscale{0.5}
\plotone{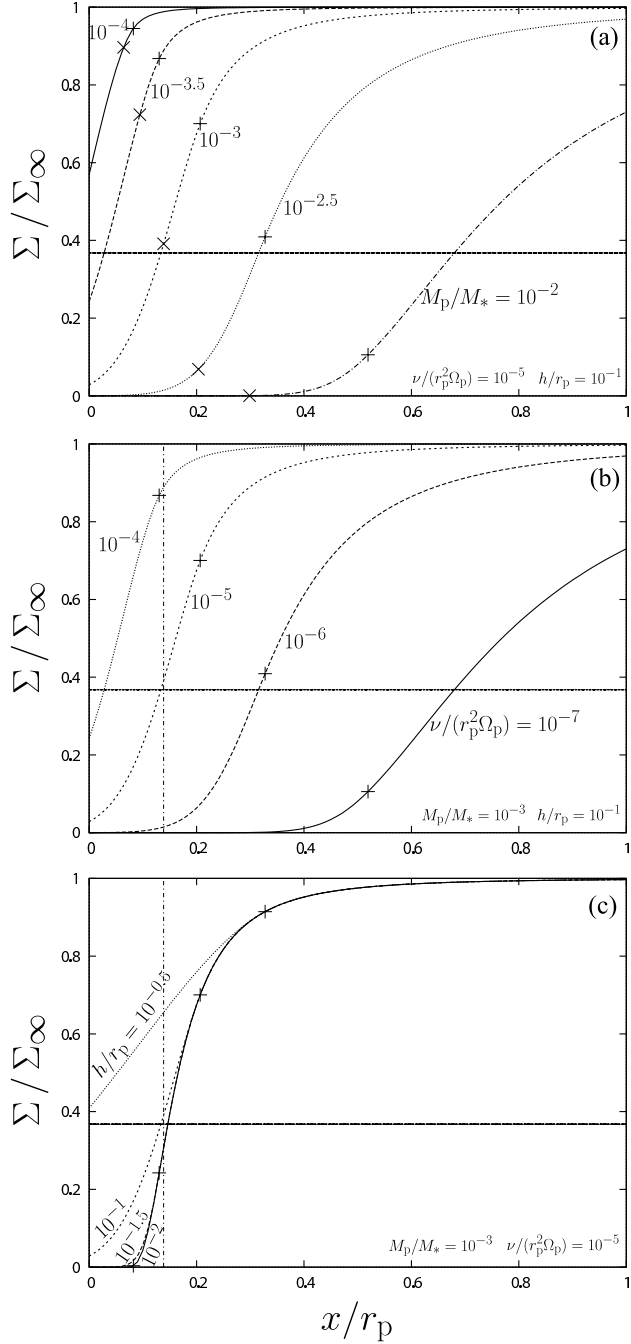}
\caption{ Surface density profiles as a function of distance from the
planetary orbit.  Top panel (a) shows planetary mass dependence ($M_{\rm
p}/M_\ast = 10^{-4}$ (solid line), $10^{-3.5}, 10^{-3}, 10^{-2.5},
10^{-2}$ (dot-dashed line)), middle panel (b) shows viscosity parameter
dependence ($\nu/(r_{\rm p}^2\Omega_{\rm p}) = 10^{-7} \mbox{(solid
line)}, 10^{-6}, 10^{-5}, 10^{-4}$(dotted line)), and bottom panel (c)
shows scale height dependence ($h/r_{\rm p} = 10^{-2} \mbox{(solid
line)}, 10^{-1.5}, 10^{-1}, 10^{-0.5}$ (dotted line)).  The standard
values of the parameters are $M_{\rm p}/M_\ast = 10^{-3}$, $\nu/(r_{\rm
p}^2 \Omega_{\rm p}) = 10^{-5}$, and $h/r_{\rm p} = 10^{-1}$.  The
position of the symbols $+$ indicates where $x=x_{\rm m}$, $\times$ in
the top panel and the vertical lines in the other two panels indicate
where $x = 2r_{\rm H}$.  $\Sigma_\infty/e$ are also shown as a
horizontal line in each panel.
\label{fig_surface_densities}}
\end{figure}

% planetary mass and viscosity dependence
Finally we briefly describe the dependence of the surface density on the
parameters.  Figure \ref{fig_surface_densities} shows the surface
density as a function of $x/r_{\rm p}$ for several different values of
planetary mass $M_{\rm p}$ (a), viscosity $\nu$ (b), and scale height
$h$ (c).  Figures \ref{fig_surface_densities}a and 2b show the gap
becomes deeper and wider with increasing planetary mass and decreasing
viscosity.
The width and depth of the gap are determined by competition between
gravitational scattering by the planet and viscous diffusion of the disk
gas.  The gap becomes wider and deeper when planet's gravity is stronger
or viscous diffusion is less efficient.  From a mathematical point of
view, this is because both $\Sigma_{\rm vis}$ and $\Sigma_{\rm R}$
depends on $M_{\rm p}$ and $\nu$ in the form of $(M_{\rm
p}^2/\nu)^{1/3}$ through $\ell$.
% scale height dependence
As for the dependence of $\Sigma$ on $h$ (see
Fig. \ref{fig_surface_densities}c), all the curves are independent of
$h$ for $x \geq x_{\rm m}$, where $\Sigma = \Sigma_{\rm vis}$ which is
independent of $h$ (see eq.[\ref{Sigma_vis}]).  In other words, surface
density in $x \geq x_{\rm m}$ is determined by the balance between
viscous torque and gravitational torque from the planet and has nothing
to do with pressure.  On the other hand, difference in surface density
appears at $x<x_{\rm m}$, where $\Sigma$ is determined by the Rayleigh
condition (see eq.[\ref{Sigma_R}]) and thus depends on all of $h$,
$\nu$, and $M_{\rm p}$ through $x_{\rm m}$.

%%%%%%%%%%%%%%%%%%%%%%%%%%%%%%%%%%%%%%%%%%%%%%%%%%%%%%%%%%%%%%%%%%%%%%%
\section{Gas accretion rate onto planets}\label{sec_accretion_rate}
%
% description of how to treat gas accretion and time evolution.
In this study, we describe the gas accretion rate onto the planet,
$\dot{M}_{\rm p}$.
Through a series of local high-resolution hydrodynamic simulations of
the gas accretion flow onto giant planets in protoplanetary disks, TW02
obtained an empirical relation between accretion rate normalized by
unperturbed surface density and $h/r_{\rm H}$, which is the only one
parameter of the local system (see eq.[18] of TW02).
Furthermore, TW02 demonstrated that only the gas in the two
bands at $|x|\sim 2r_{\rm H}$ accretes onto the planet.  Thus, writing
equation (18) of TW02 in the explicit form, for $0.5 \leq
h/r_{\rm H} \leq 1.8$, we have
\begin{equation}
\dot{M}_{\rm p}
 = \dot{A} \Sigma_{\rm acc},
\label{mdot}
\end{equation}
where
\begin{eqnarray}
\dot{A}
%&=& 1.26 \kakkoi{h}{r_{\rm H}}{-2.0}
%         \kakkoi{r_{\rm H}}{a}{2} \\
&\simeq&
  0.29
  \kakkoi{h}{r_{\rm p}}{-2}
  \kakkoi{M_{\rm p}}{M_\ast}{4/3}
  r_{\rm p}^2 \Omega_{\rm p},
\label{Adot}
\end{eqnarray}
which is the area, the gas in which is to be accreted onto the planet
per unit time (hereafter the {\it accretion area}), and $\Sigma_{\rm
acc} = \Sigma(2r_{\rm H})$.

The reason why small $h$ or large $M_{\rm p}$ yields high $\dot{M}_{\rm
p}$ is explained as follows.  Disk gas loses its energy by passing
through spiral shocks around the planet and consequently accretes onto
the planet.  Small $h$ (i.e., small sound velocity) or large $M_{\rm p}$
(i.e., strong perturbation on gas motion) yields a large Mach number of
the disk gas, which results in strong shocks.  In addition, an increase
in $M_{\rm p}$ expands the planetary feeding zone, resulting in large
$\dot{M}_{\rm p}$.

Note that the position of the accretion band, $x_{\rm acc}$, is located
at $\sim 2r_{\rm H}$ in the highest-mass case of TW02 (i.e.,
$\tilde{C}_{\rm iso}=0.5$ in their notation; see Fig. 7 of TW02 where
$x_{\rm acc}$ is shown to depend slightly on $h/r_{\rm H}$).  When
$\tilde{C}_{\rm iso} < 0.5$, a gap exists around the planetary orbit.
In such situations, appropriate choice of $x_{\rm acc}$ is crucial in
determining $\dot{M}_{\rm p}$ because a small difference in $x_{\rm
acc}$ makes a large difference in $\Sigma_{\rm acc}$ and thus
$\dot{M}_{\rm p}$, while choice of $x_{\rm acc}$ has little influence on
$\dot{M}_{\rm p}$ in low-mass cases.  That is why we adopt the value of
$x_{\rm acc} (=2r_{\rm H})$ for the highest-mass case of TW02.

Note also that, as shown by \citet{DKH03}, the gas accretion rate
obtained by two-dimensional simulations are usually higher than that
obtained by three-dimensional simulations.  Thus the accretion rate
given by equation (\ref{mdot}) could be overestimated.  We will discuss
this issue in \S \ref{sec_remarks}.

%%%%%%%%%%%%%%%%%%%%%%%%%%%%%%%%%%%%%%%%%%%%%%%%%%%%%%%%%%%%%%%%%%%%%%%
\section{Evolution}
In this section we show the evolution of the growth rate and mass of an
accreting planet.  To gain a proper understanding of the basic behavior
of the planetary accretion, we first explore two simple cases with no
disk dissipation (i.e. constant $\Sigma_\infty$) and with
exponentially-decreasing $\Sigma_\infty$ in \S \S
\ref{sec_without_dissipation} and \ref{sec_with_dissipation},
respectively.  Then, we investigate the accretion of a giant planet
embedded in a viscously evolving protoplanetary disk in \S
\ref{sec_self-similar}.

The numerical procedure is as follows.  Except in \S
\ref{sec_without_dissipation}, we first calculate the unperturbed
surface density, $\Sigma_\infty$, from equation~(\ref{Sigma_infty_init})
in \S \ref{sec_with_dissipation} or from equation~(\ref{Sigma_infty_ss})
in \S \ref{sec_self-similar}.  For given values of the parameters,
$h/r_{\rm p}$ and $\nu/(r_{\rm p}^2\Omega_{\rm p})$, we then calculate
$\Sigma_{\rm acc}$ and $\dot{A}$, using equation (\ref{Sigma_vis}) if
$x_{\rm acc} \geq x_{\rm m}$ or equation (\ref{Sigma_R}) if $x_{\rm acc}
\leq x_{\rm m}$. Finally we integrate $\dot{M}_p$ with respect to time
using $\Sigma_{\rm acc}$ and $\dot{A}$ (see eq.~[\ref{mdot}]) to obtain
the time evolution of the planetary mass.

%======================================================================
\subsection{Case without disk-dissipation}\label{sec_without_dissipation}
We first show the evolution of the planetary mass without global disk
dissipation. In this case the input parameters are
$\Sigma_\infty r_{\rm p}^2/M_\ast$, $h/r_{\rm p}$, and $\nu/(r_{\rm
p}^2\Omega_{\rm p})$.
% Since
% protoplanetary disks dissipate on a finite timescale, the assumption
%of no dissipation is unrealistic.  However this kind of simple
%experiment helps us to understand the basic behavior of the evolution
%of the planetary mass.

%----------------------------------------------------------------------
\subsubsection{General Properties}
\begin{figure}
\epsscale{1.0}
\plotone{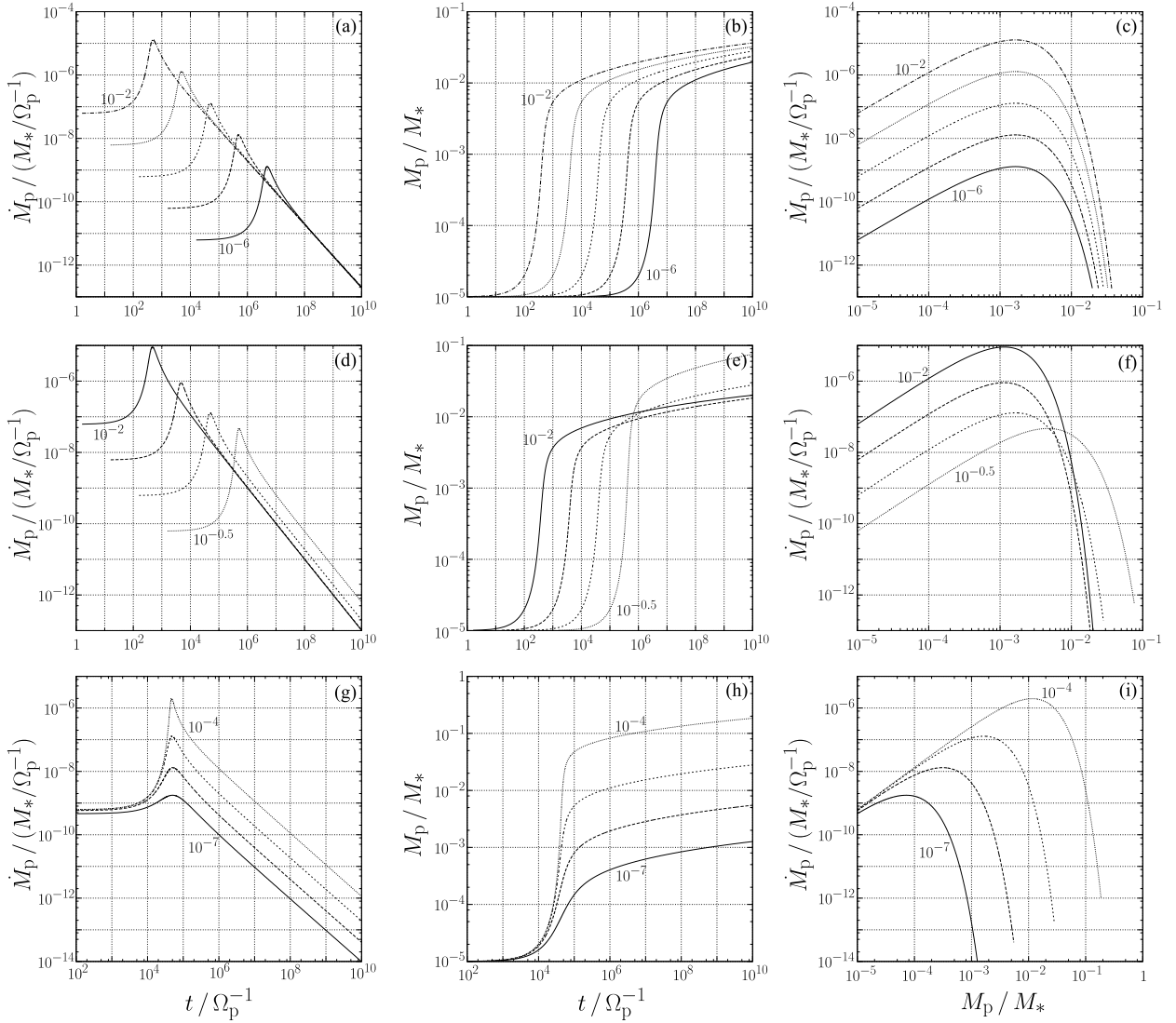}
\caption{Evolution of planetary mass and accretion rate without disk
dissipation.  Figures in the left column show accretion rate onto the
planet as a function of time, those in the central column show planetary
mass as a function of time, and those in the right column show accretion
rate vs planetary mass.  Figures in the top row show the dependence on
surface density ($\Sigma_\infty/(M_\ast r_{\rm p}^{-2}) = 10^{-6}
\mbox{(solid line)}, 10^{-5}, 10^{-4}, 10^{-3}, 10^{-2}$ (dot-dashed
line)), those in the middle row show the dependence on scale height
($h/r_{\rm p} = 10^{-2} \mbox{(solid line)}, 10^{-1.5}, 10^{-1},
10^{-0.5}$ (dotted line)), and those in the bottom row show the
dependence on viscosity coefficient ($\nu/(r_{\rm p}^2\Omega_{\rm p}) =
10^{-7} \mbox{(solid line)}, 10^{-6}, 10^{-5}, 10^{-4}$ (dotted line)).
The standard values of the parameters are $\Sigma_\infty/(M_\ast r_{\rm
p}^{-2}) = 10^{-4}$, $\nu/(r_{\rm p}^2 \Omega_{\rm p}) = 10^{-5}$, and
$h/r_{\rm p} = 10^{-1}$.
\label{fig_nine_figures}}
\end{figure}

% general explanation of evolution
Figure \ref{fig_nine_figures} shows the evolution of the planetary mass
and the gas accretion rate for several values of the three parameters.
Without global disk depletion, the only way to suppress gas accretion is
opening a deep gap around the planetary orbit by the planet itself.
As seen in Fig.~\ref{fig_nine_figures}, the evolution can be divided
into two phases.  The first phase is growth without a gap (pre-gap
phase), while the second phase is growth with a deep gap (post-gap
phase).  Properties of the evolution in both phases can be explained in
analytical ways below.

In the pre-gap phase, $\dot{M}_{\rm p}$ is almost proportional to
$M_{\rm p}^{4/3}$ (see Figs.~\ref{fig_nine_figures}--(c),(f),(i)).  In
this phase, the planetary mass is insufficient to open a gap, so that
$\Sigma_{\rm acc} \simeq \Sigma_\infty$.  Then we can easily integrate
equation~(\ref{mdot}) with equation (\ref{Adot}):
\begin{equation}
\dot{M}_{\rm p}
 = \frac
   {\displaystyle
    S \Sigma_\infty r_{\rm p}^2 \Omega_{\rm p}
    \kakkoi{h}{r_{\rm p}}{-2}}
   {\displaystyle
    \left[ \kakkoi{M_{\rm p,init}}{M_\ast}{-1/3}
	  -\frac{S}{3}
	   \kakkoi{\Sigma_\infty}{M_\ast / r_{\rm p}^2}{}
	   \kakkoi{h}{r_{\rm p}}{-2}
	   \kakkoi{t}{\Omega_{\rm p}^{-1}}{}
    \right]^4},
\end{equation}
where $S$ corresponds to 0.29 in equation~(\ref{Adot}) and $M_{\rm
p,init}$ is the initial mass of the planet that
corresponds to the mass beyond which the gas accretion is limited by
disk-gas supply.  This equation implies that if $\Sigma_{\rm acc}$ is
constant, the gas accretion rate diverges at a time
\begin{eqnarray}
\tau_{\rm div}
&=&
 \frac{3}{S}
 \kakkoi{M_{\rm p,init}}{M_\ast}{-1/3}
 \kakkoi{\Sigma_\infty}{M_\ast / r_{\rm p}^2}{-1}
 \kakkoi{h}{r_{\rm p}}{2}
 \Omega_{\rm p}^{-1} \nonumber \\
&=&
 4.8 \times 10^4
 \kakkoi{M_{\rm p,init}}{10^{-5}M_\ast}{-1/3}
 \kakkoi{\Sigma_\infty}{10^{-5}M_\ast / r_{\rm p}^2}{-1}
 \kakkoi{h}{10^{-1.5}r_{\rm p}}{2}
 \Omega_{\rm p}^{-1};
\label{tau_div}
\end{eqnarray}
$\tau_{\rm div}$ corresponds to the end of the pre-gap phase.
In the post-gap phase, $\dot{M}_{\rm p}$ decreases with $M_{\rm p}$ in
an exponential fashion (see Figs.~\ref{fig_nine_figures}--(c),(f),(i)).
Such a steep decrease in $\dot{M}_{\rm p}$ is due to a steep decrease in
$\Sigma_{\rm acc}$ with respect to $M_{\rm p}$ via $\ell$, $x_{\rm m}$,
and $r_{\rm H}$ (see eqs.[\ref{Sigma_vis}] and [\ref{Sigma_R}]).  We can
derive approximate expressions for the accretion rate in this phase,
which shows $\dot{M}_{\rm p}$ is inversely proportional to time as shown
in Figs.~\ref{fig_nine_figures}(a)-(c) (see appendix for the detail).

%----------------------------------------------------------------------
\subsubsection{Dependence on the parameters}
% sigma dependence
We now see the dependence of the evolution of planetary mass on the
unperturbed surface density, $\Sigma_\infty$
(Figs.~\ref{fig_nine_figures}--(a),(b),(c)).
The evolution timescale is inversely proportional to $\Sigma_\infty$
(e.g., Fig. \ref{fig_nine_figures}-(b)).  This is simply because
$\Sigma_{\rm acc}$ is proportional to $\Sigma_\infty$, whereas $\dot{A}$
is independent of $\Sigma_\infty$.  The peak value of $\dot{M}_{\rm p}$
is thus proportional to $\Sigma_\infty$ accordingly (e.g.,
Fig. \ref{fig_nine_figures}-(c)).
%

% scale height dependence
The dependence on scale height, $h$, is shown in
Figs.~\ref{fig_nine_figures}--(d),(e),(f).  In the pre-gap phase,
$\dot{M}_{\rm p}$ is simply proportional to $h^{-2}$ and the evolution
timescale is thus $\propto h^2$.  This is because the accretion area
$\dot{A}$ $\propto h^{-2}$ (see eq.[\ref{Adot}]) in the pre-gap phase
where $\Sigma_{\rm acc}=\Sigma_\infty$.  Even in the post-gap phase, the
dependence of $\dot{M}_{\rm p}$ on $h$ is the same as that in the
pre-gap phase if $h$ (i.e., $x_{\rm m}$) is small ($h/r\lesssim 0.03$).
As long as $h$ is so small that $x_{\rm m} \leq 2r_{\rm H}$,
$\Sigma_{\rm acc}$ is determined by $\Sigma_{\rm vis}$ and is thus
independent of $h$ (see eqs.[\ref{Sigma_vis}] and
[\ref{Sigma_general}]).
When $h$ is so large that $x_{\rm m} \geq 2r_{\rm H}$, $\Sigma_{\rm acc}
= \Sigma_{\rm R}$, which depends on $h$.  Hence $\Sigma_{\rm acc}$
increases with $h$ for large $h$.
The transition between the low-$h$ and high-$h$ cases takes place when
$x_{\rm m} = 2r_{\rm H}$, namely, from equation (\ref{x_m_typical}):
\begin{equation}
h
 = 0.037 r_{\rm p}
   \kakkoi{M_{\rm p}}{10^{-3}M_\ast}{-1/6}
   \kakkoi{\nu}{10^{-5}r_{\rm p}^2 \Omega_{\rm p}}{1/2}.
\end{equation}

% viscosity dependence 
The dependence on viscosity, $\nu$, is shown in
Fig.\ref{fig_nine_figures}-(g)(h)(i).
In the pre-gap phase, the evolution of $\dot{M}_{\rm p}$ is almost the
same for different values of $\nu$, because $\Sigma_{\rm acc} $ is
almost constant ($\simeq \Sigma_\infty$) in the pre-gap phase and
$\dot{A}$ is independent of $\nu$.  Thus the timing when $\dot{M}_{\rm
p}$ reaches at the peak value does not depend on $\nu$ either (see
eq.[\ref{tau_div}]).
In the post-gap phase, $\dot{M}_{\rm p}$ at a given $t$ increases with
$\nu$.  In appendix \ref{Appendix-B}, we have derived approximate
solutions for $\dot{M}_{\rm p}$ in the post-gap phase.  From equations
(\ref{M_trans}) and (\ref{mdot_analytic_1}), one finds that $M_{\rm p}
\propto \nu$ in high-$\nu$ cases (i.e., $x_{\rm acc} > x_{\rm m}$).  In
low-$\nu$ cases (i.e., $x_{\rm acc} < x_{\rm m}$), $\dot{M}_{\rm p}$
also increases with $\nu$, but the dependence is rather weak (see
eqs.[\ref{dfdMp}] and [\ref{mdot_analytic_2}]).
Also, the accretion rate is inversely proportional to time (see
eqs.[\ref{mdot_analytic_1}] or [\ref{mdot_analytic_2}]), so that the
maximum accretion rate can be roughly estimated by
eqs.[\ref{mdot_analytic_1}] or [\ref{mdot_analytic_2}] at $t=\tau_{\rm
div}$.

%======================================================================
\subsection{Case with exponential disk dissipation}\label{sec_with_dissipation}
Next we examine the evolution of the planetary mass in a simple case
where the disk surface density decreases in an exponential fashion with
a time constant of $\tau_{\rm dep}$:
\begin{equation}
\Sigma_\infty
 = \Sigma_{\infty,\rm init} \exp \left(-\frac{t}{\tau_{\rm dep}}\right),
 \label{Sigma_infty_init}
\end{equation}
where $\Sigma_{\infty,\rm init}$ is the initial surface density at
infinity.

\begin{figure}
\epsscale{1.0}
\plotone{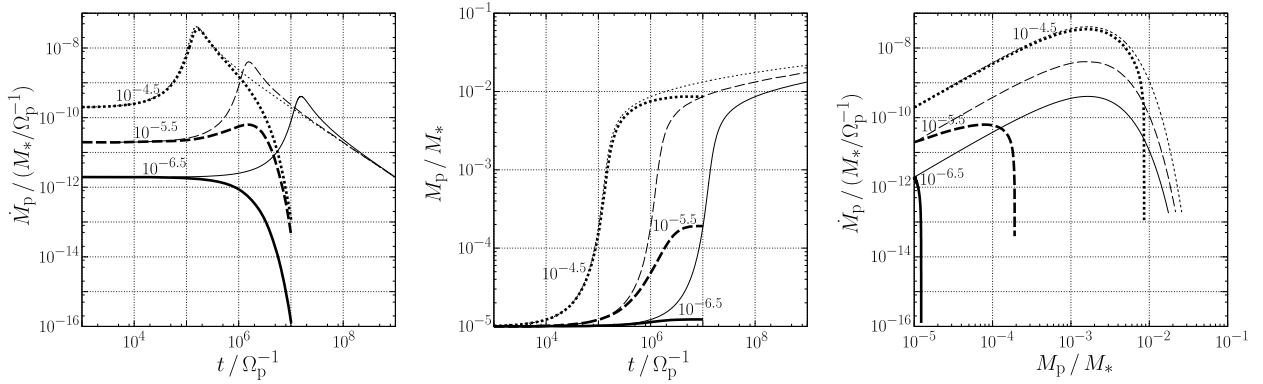}
\caption{Evolution of planetary mass and accretion rate with disk
dissipation when $h/r_{\rm p}=0.1$, $\nu/(r_{\rm p}^2\Omega_{\rm
p})=10^{-5}$.  Three panels are the same axis with
Fig.~\ref{fig_nine_figures}.  Three thick lines show different initial
surface densities ($\Sigma_{\rm init}/(M_\ast r_{\rm p}^{-2}) = 10^{-6.5}
$(solid line), $10^{-5.5}$ (long-dashed line), $10^{-4.5}$
(short-dashed line)) and evolutions without disk dissipation are also
shown as thin lines.
\label{fig_t-m-mdot_mixed}}
\end{figure}

% Explanation of fig.4
Figure \ref{fig_t-m-mdot_mixed} shows the evolution of $M_{\rm p}$
and $\dot{M}_{\rm p}$ for $\tau_{\rm dep} = 10^6 \Omega_{\rm
p}^{-1}$ and three values of $\Sigma_{\infty,\rm init}$.
In the high-$\Sigma_{\infty,\rm init}$ case ($\Sigma_{\infty,\rm init}/
(M_\ast r_{\rm p}^{-2})=10^{-4.5}$), $\dot{M}_{\rm p}$ increases with
time to reach a peak at $t\sim1.5\times 10^5 \Omega_{\rm p}^{-1}$ and
then decreases with time, which is similar to the evolution without disk
dissipation shown in \S \ref{sec_without_dissipation}. This is because
$\tau_{\rm div} < \tau_{\rm dep}$ in this case.
In the low-$\Sigma_{\infty,\rm init}$ case ($\Sigma_{\infty,\rm
init}/(M_\ast r_{\rm p}^{-2}) = 10^{-6.5}$), on the other hand,
$\dot{M}_{\rm p}$ decreases without experiencing a significant increase
that occurs in the high-$\Sigma_{\infty,\rm init}$ case.  This is
because $\tau_{\rm div} > \tau_{\rm dep}$ in this case
($\tau_{\rm div} \propto \Sigma_{\infty,\rm init}^{-1}$; see
eq.[\ref{tau_div}]).
Because of such different evolution of $\dot{M}_{\rm p}$, the mass
evolution also differs between high- and low-$\Sigma_{\infty,\rm init}$
cases (see Fig.~\ref{fig_t-m-mdot_mixed}b): The planet captures a
significant amount of gas to be a Jupiter-like planet in the
high-$\Sigma_{\infty,\rm init}$ case, whereas the planet captures only a
small amount of gas to be a Neptune-like planet in the
low-$\Sigma_{\infty,\rm init}$ case.  The boundary between the two
regimes is determined by the condition $\tau_{\rm dep} = \tau_{\rm
div}$, namely, from equation (\ref{tau_div}),
\begin{equation}
\frac{\Sigma_{\infty, \rm init}}{M_\ast / r_{\rm p}^2}
 = \frac{3}{S}
   \kakkoi{M_{\rm p,init}}{M_\ast}{-1/3}
   \kakkoi{h}{r_{\rm p}}{2}
   \kakkoi{\tau_{\rm dep}}{\Omega_{\rm p}^{-1}}{-1}.
\end{equation}

% Introducing ``zeta''
%We next consider about the relationship between $\Sigma_{\infty,\rm
%init}$ and $\tau_{\rm dep}$.  Substituting $t=\tilde{t} \tau_{\rm dep}$,
%$M_{\rm p} = \tilde{M}_{\rm p}M_\ast$, and $\dot{A} = \dot{\tilde{A}}
%r_{\rm p}^2 \Omega_{\rm p}$ into equation (\ref{mdot}), we have
%\begin{equation}
%\frac{d\tilde{M}_{\rm p}}{d\tilde{t}}
% = \dot{\tilde{A}} f_{\rm acc} \exp\left(-\tilde{t}\right)
%   \kakkoi{\Sigma_{\infty,\rm init}}{M_\ast/r_{\rm p}^2}{}
%   \kakkoi{\tau_{\rm dep}}{\Omega_{\rm p}^{-1}}{},
%\label{mdot_tilde}
%\end{equation}
%where $f_{\rm acc} \equiv \Sigma_{\rm acc}/\Sigma_\infty$.  Since both
%$\dot{\tilde{A}}$ and $f_{\rm acc}$ are independent of both
%$\Sigma_{\infty, \rm init}$ and $\tau_{\rm dep}$, the evolution of the
%normalized planetary mass $\tilde{M}_{\rm p}$ depends via the product of
%$\Sigma_{\infty, \rm init}$ and $\tau_{\rm dep}$, not on the two
%parameters independently; the final mass also depends through the
%product of the two parameters accordingly.  We hence define
%\begin{equation}
%\zeta
% \equiv
%  \kakkoi{\Sigma_{\infty, \rm init}}{M_\ast/r_{\rm p}^2}{}
%  \kakkoi{\tau_{\rm dep}}{\Omega_{\rm p}^{-1}}{},
%\label{zeta}
%\end{equation}
%as a new parameter of our calculation.

\begin{figure}
\epsscale{0.8}
\plotone{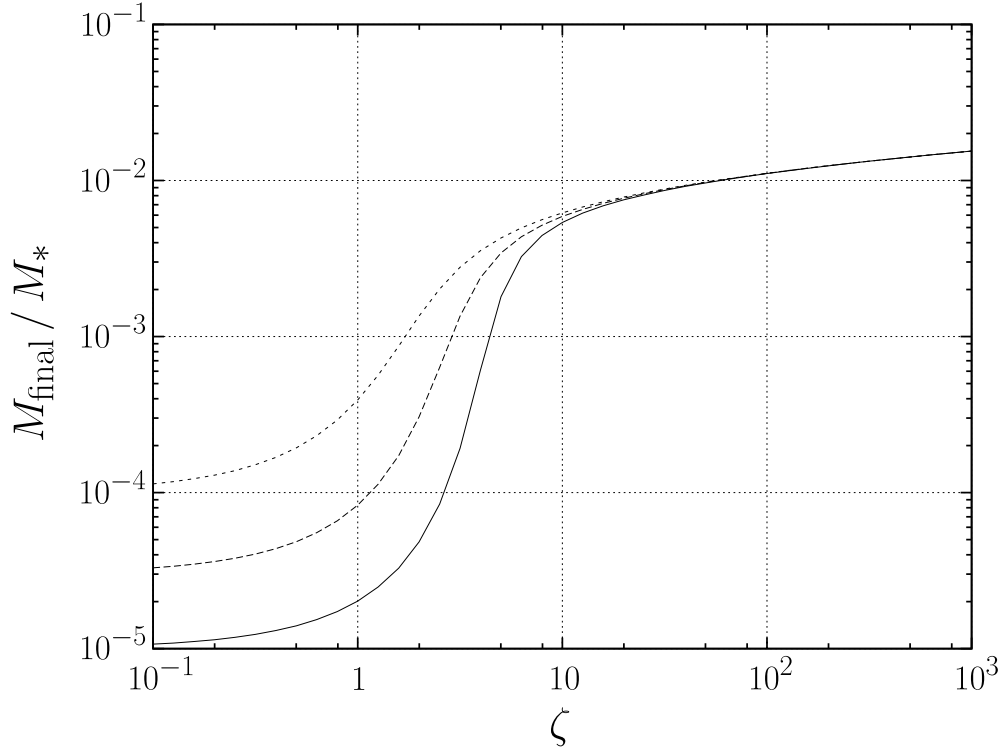}
\caption{Final mass of planets as a function of $\zeta$ (defined at
eq.[\ref{zeta}]) in the case with $h/r_{\rm p}=0.1$, $\nu/ (r_{\rm
p}^2\Omega_{\rm p})=10^{-5}$.  The three lines are different initial
planetary mass ($M_{\rm p, init}/M_\ast = 1\times 10^{-5}, 3\times
10^{-5}, 1\times 10^{-4}$).
\label{fig_final_mass_vs_zeta}}
\end{figure}

% Explanation of Fig.5
Figure \ref{fig_final_mass_vs_zeta} shows the final mass defined by
\begin{equation}
M_{\rm final}
\equiv
 \int_0^\infty
 \dot{M}_{\rm p} dt
 \label{M_final}
\end{equation}
as a function of a quantity $\zeta$ defined by
\begin{equation}
\zeta
 \equiv
  \kakkoi{\Sigma_{\infty, \rm init}}{M_\ast/r_{\rm p}^2}{}
  \kakkoi{\tau_{\rm dep}}{\Omega_{\rm p}^{-1}}{}.
\label{zeta}
\end{equation}
The dependence of $M_{\rm final}$ on $\tau_{\rm dep}$ is the same as
that on $\Sigma_{\infty,\rm init}$ because the ratio of $\tau_{\rm div}$
($\propto \Sigma_{\infty,\rm init}^{-1}$) to $\tau_{\rm dep}$ determines
the evolution.
Figure \ref{fig_final_mass_vs_zeta} illustrates that when $\zeta$ is
small ($\lesssim 1$), the final mass is almost the same as the initial
mass.  In this regime, the planet captures only a tiny amount of gas and
becomes a Neptune-like planet.
When $\zeta \geq 10$, on the other hand, the planet captures substantial
gas to be a gas giant planet like Jupiter and the final mass does not
depend on the initial planetary mass.  This is because the planet
becomes large enough to form a gap and the evolution slows down
significantly after gap formation; the condition for gap opening is
determined by the planetary mass.
The transition occurs when $\tau_{\rm div}=\tau_{\rm dep}$, namely, from
equation (\ref{tau_div}),
\begin{equation}
\zeta_{\rm t}
 \simeq 4.8
        \kakkoi{M_{\rm p,init}}{10^{-5}M_\ast}{-1/3}
        \kakkoi{h}{0.1r_{\rm p}}{2}.
\label{zeta_t}
\end{equation}

%%%
Figure \ref{fig_final_mass_contour} shows that the final mass as a
function of $\zeta$ and $\nu$ in the cases with $h/r=0.1$ and
$h/r=0.032$.
A portion characterized by a steep gradient is illustrated to shift to
the left with decreasing $h/r_{\rm p}$ ($\zeta_{\rm t} \sim 5$ for
$h/r_{\rm p}=0.1$, while $\zeta_{\rm t} \sim 0.1$ for $h/r_{\rm
p}=0.032$, see eq.[\ref{zeta_t}]).
On the left side to the portion (low-$\zeta$ case), the final mass is
almost equal to the initial mass and thus almost independent of
viscosity.
On the right side, the final mass increases with viscosity.  This is
because the growth is limited by opening a deep gap; the mass for gap
opening depends on viscosity.

%======================================================================
\subsection{Case with global disk evolution}
\label{sec_self-similar}
\subsubsection{the model} \label{sec_model}
We have so far neglected global evolution of the disk, although we
examined a simple case with an exponential depletion of the disk gas in
\S \ref{sec_with_dissipation}.  In reality, the unperturbed surface
density, $\Sigma_\infty$, changes because of global disk evolution.  In
this section, we examine the effects of global disk evolution on
planetary growth.
% (A)

Because a detailed description of disk evolution is beyond the scope of
this paper, we assume that the disk surface density changes with time in
such a way
\begin{equation}
\Sigma_\infty
 = \Sigma_{\rm ss}(r,t)
   \exp\left(-\frac{t}{\tau_{\rm dep}}\right),
\label{Sigma_infty_ss}
\end{equation}
instead of equation (\ref{Sigma_infty_init}): $\Sigma_{\rm ss}$
represents a change in the surface density due to viscous diffusion of
the disk gas and $e^{-t/\tau_{\rm dep}}$ is introduced to mimic
photoevaporation of the disk.  For $\Sigma_{\rm ss}$, we adopt the
self-similar solution with $\alpha$ prescription for disk viscosity
given by \citep{Hartmann98}
\begin{equation}
\Sigma_{\rm ss}(r,t)
 = \frac{M_{\rm disk}}{2\pi R_{\rm out}^2}
   \kakkoi{r}{R_{\rm out}}{-1}
   \tilde{\tau}_{\rm ss}^{-3/2}
   \exp \left(-\frac{r/R_{\rm out}}{\tilde{\tau}_{\rm ss}}\right),
   \label{Sigma_self_similar}
\end{equation}
where $M_{\rm disk}$ is the initial disk mass and $R_{\rm out}$ is the
initial disk size; $\tilde{\tau}_{\rm ss}$ is defined as
\begin{equation}
\tilde{\tau}_{\rm ss}
 = \frac{t}{\tau_{\rm vis}} + 1,
\label{tilde_tau_ss}
\end{equation}
with a typical timescale of global viscous evolution
\begin{eqnarray} 
\tau_{\rm vis} &\equiv& \frac{R_{\rm out}^2}{3\nu_{\rm out}} \nonumber \\
               &=& 5.3 \times 10^5
               \kakkoi{\alpha}{0.01}{-1}
               \kakkoi{h_{\rm 1AU}}{10^{-1.5}\rm AU}{-2}
               \kakkoi{R_{\rm out}}{\rm 100AU}{}
               {\rm yr},
\label{tau_vis}
\end{eqnarray}
where $\nu_{\rm out}$ is viscous coefficient of the disk gas at
$r=R_{\rm out}$.  In deriving equation (\ref{Sigma_self_similar}), we
have assumed a temperature distribution, $T\propto r^{-1/2}$,
% and then calculated $\nu$ from $\nu=\alpha c h$, 
which results in $\nu \propto r$.
%

%%%
%We now calculate gas accretion rate onto planets.
%
Using the equations above, we calculate the \textit{local} gas accretion
rate onto the planet, denoted hereafter by $\dot{M}_{\rm p,local}$, in
the same way as we did in \S \ref{sec_with_dissipation}.
However, when $\dot{M}_{\rm p,local}$ is larger than the radial mass
flux due to viscous diffusion in the disk, the latter limits the
planetary growth.
The mass flux through a ring with radius $r$ is given by
\begin{eqnarray}
\dot{M}_{\rm disk}(r,t)
&=& 2\pi r \Sigma_{\rm ss}(r,t) v_r \nonumber \\
&=& \frac{M_{\rm disk}}{\tau_{\rm vis}}
    \left( \frac{1}{2} - \frac{r}{\tilde{\tau}_{\rm ss}R_{\rm out}}\right)
    \tilde{\tau}_{\rm ss}^{-3/2}
    \exp \left(-\frac{r}{\tilde{\tau}_{\rm ss}R_{\rm out}}\right),
 \label{flux_self_similar}
\end{eqnarray}
with the radial drift velocity of the diffusing gas in a Keplerian disk
\begin{equation}
v_r
 = -\frac{3\nu}{r}
    \left( \pfrac{\ln(\Sigma\nu)}{\ln r} + \frac{1}{2}\right).
\end{equation}
Thus we calculate the gas accretion rate onto the planet in such a way
\begin{equation}
\dot{M}_{\rm p}
=\left\{
 \begin{array}{ll}
  \dot{M}_{\rm p,local} & \mbox{if}\ 
  \dot{M}_{\rm p,local} < \dot{M}_{\rm disk} \\
  \dot{M}_{\rm disk} & \mbox{if}\ 
  \dot{M}_{\rm p,local} > \dot{M}_{\rm disk}.
 \end{array}
 \right.
\label{Mdot_p_min}
\end{equation}
%\begin{equation}
%\dot{M}_{\rm p}
%=\left\{
% \begin{array}{ll}
%  \dot{M}_{\rm p,ss} & \mbox{if}\ \dot{M}_{\rm p,ss} < \dot{M}_{\rm disk} \\
%  \dot{M}_{\rm disk} & \mbox{if}\ \dot{M}_{\rm p,ss} > \dot{M}_{\rm disk}.
% \end{array}
% \right.
%\label{Mdot_p_min}
%\end{equation}

%%%%%%%% Comments on previous studies relevant to gas accretion rate
% We should note here that 
There are several studies that modeled gas accretion rate onto gas giant
planets.
\citet{Guillot06} used a formula in which the gas accretion rate onto
the planet is 0.3 times the radial mass flux of the disk gas due to
global viscous diffusion.  However their formula is an empirical one and
applies in a limited situation.
\citet{LSA99} carried out a series of hydrodynamic simulations to obtain
gas accretion rate onto planets.  Their simulations demonstrated that
the planetary accretion rate can be larger than the diffusion flux
because of gradients imposed by the gap.  However, what they found is
probably a transient phenomenon that occurs on a timescale much shorter
than the viscous timescale.
\citet{VA04} used an approximate formula based on hydrodynamic
simulations done by \citet{LSA99} and \citet{DHK02}.
%Their gas accretion rate depends on planetary mass as $M_{\rm p}^{1/3}$
%times $\exp(-M_{\rm p})$; the former may correspond to a geometrical
%factor proportional to the Hill radius and the latter corresponds to gas
%depletion due to gap formation.  But the geometrical factor should be
%proportional to the area or volume whose size is the Hill radius;
%$M_{\rm p}^{2/3}$ or $M_{\rm p}^1$ should be better in the sense.  Their
%formula, furthermore, includes no dependence on disk temperature.

%%%%%%%%%%%%%%%%%%%%
%%% results
%
\subsubsection{dependence on the disk parameters}\label{sec:dependence on the disk parameters}

\begin{figure}
\epsscale{1.0}
\plottwo{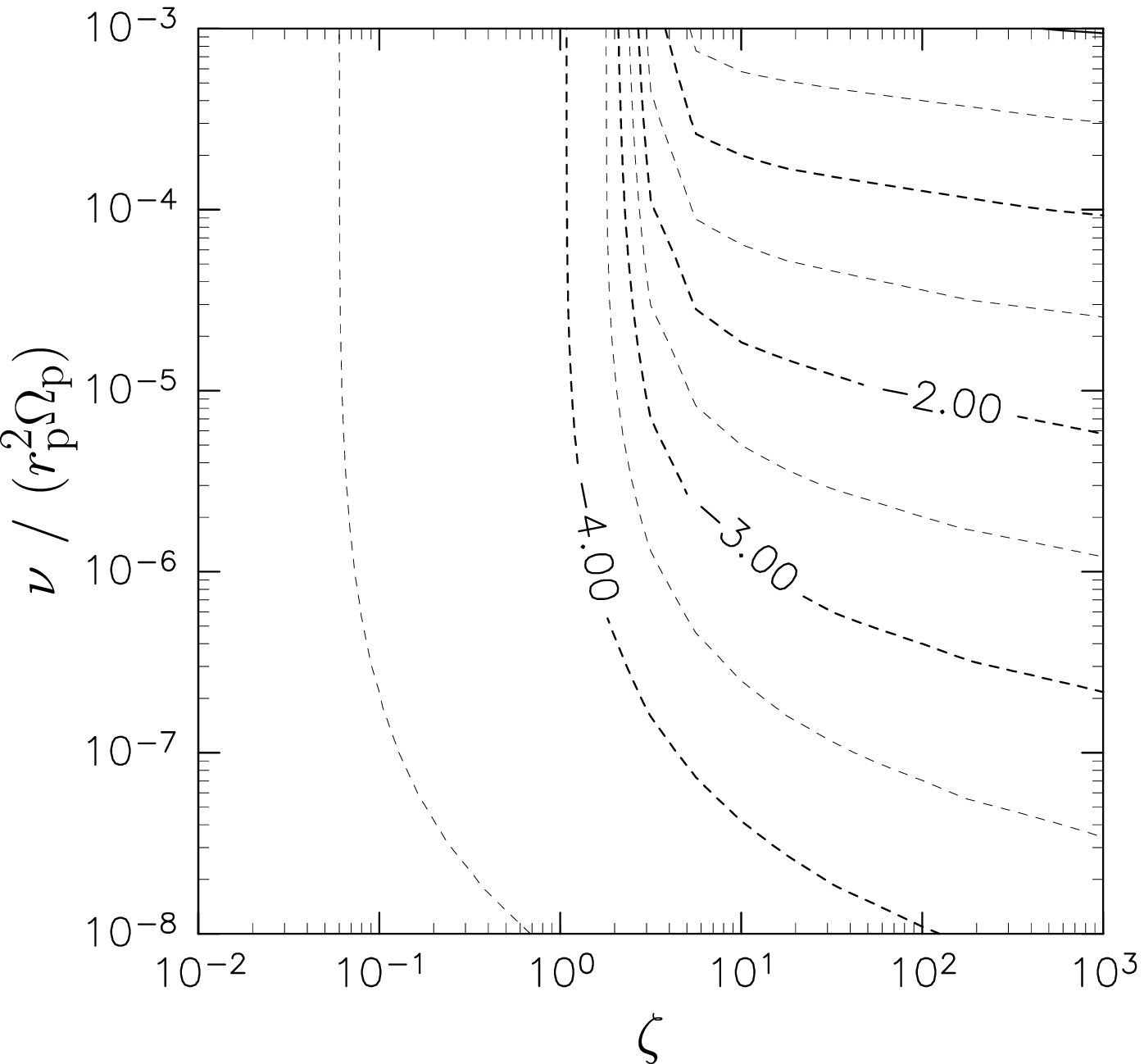}{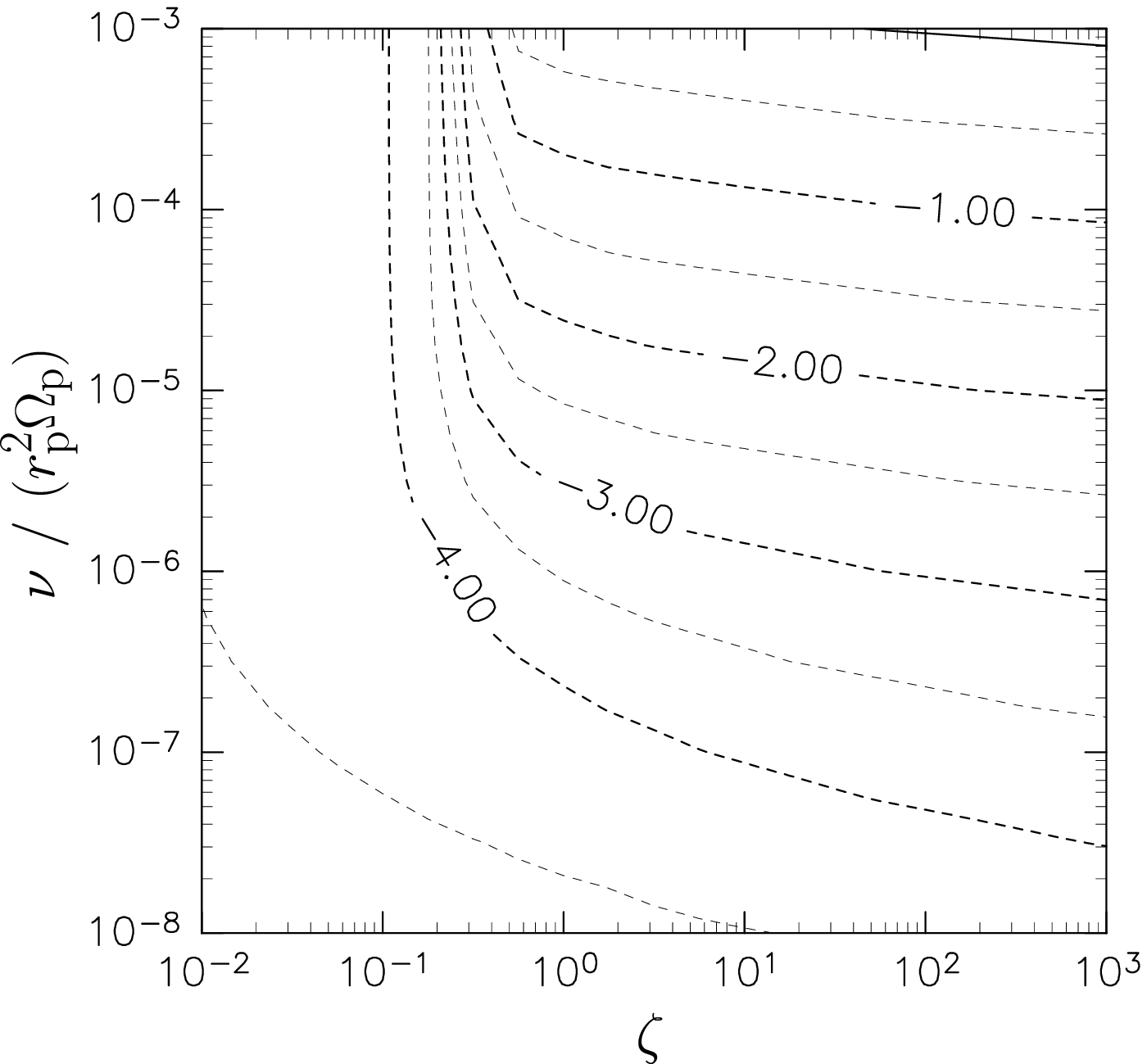}
\caption{Contour lines of $\log(M_{\rm final}/M_\ast)$ as a function
of $\zeta$ and $\nu/(r_{\rm p}^2\Omega_{\rm p})$.  Left panel is when
$h=0.1r_{\rm p}$, right panel is when $h=10^{-1.5}r_{\rm p}$.  The
initial mass is set to $10^{-5}M_\ast$.
\label{fig_final_mass_contour}}
\end{figure}

\begin{figure}
\epsscale{1.0}
\plottwo{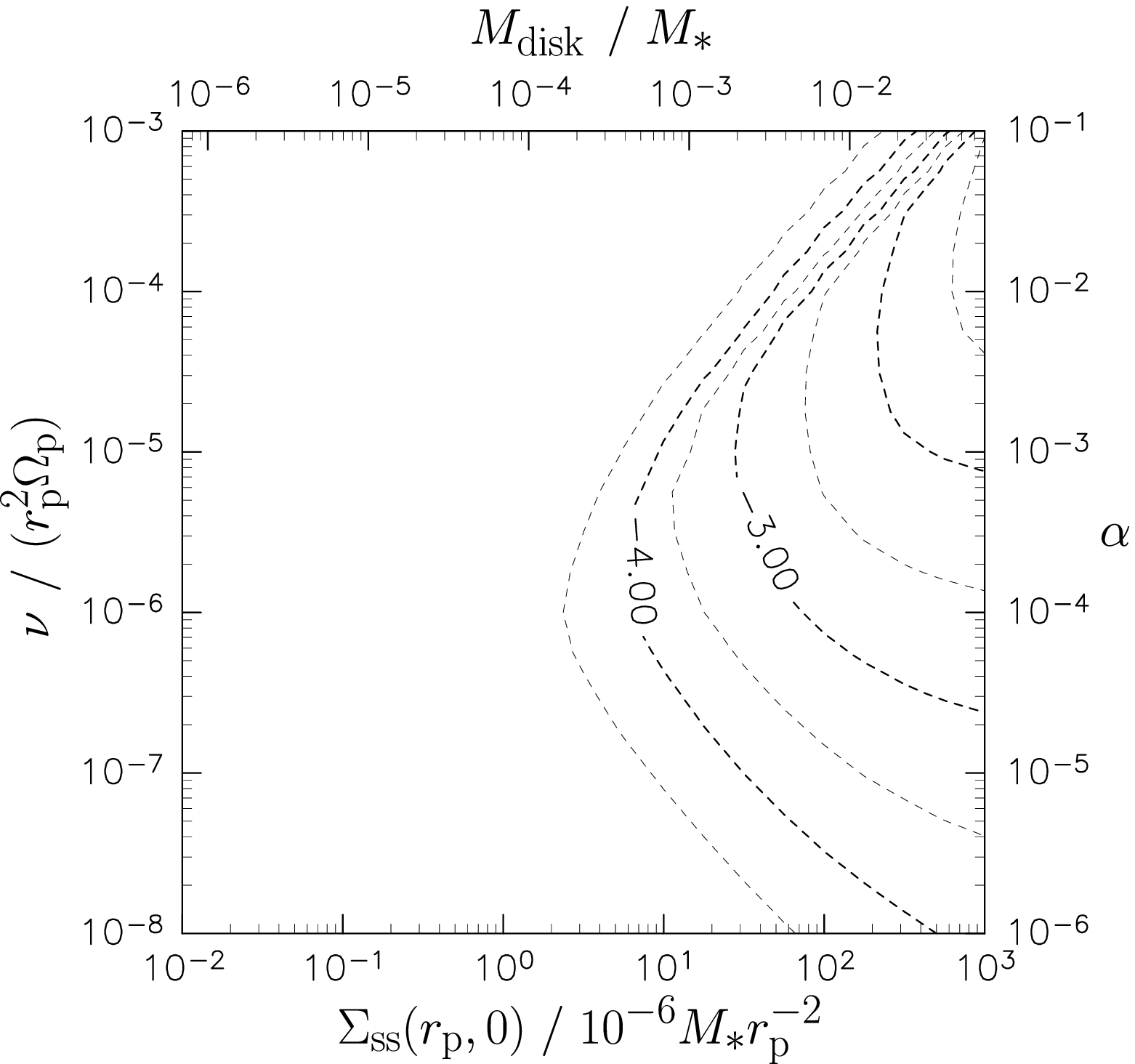}{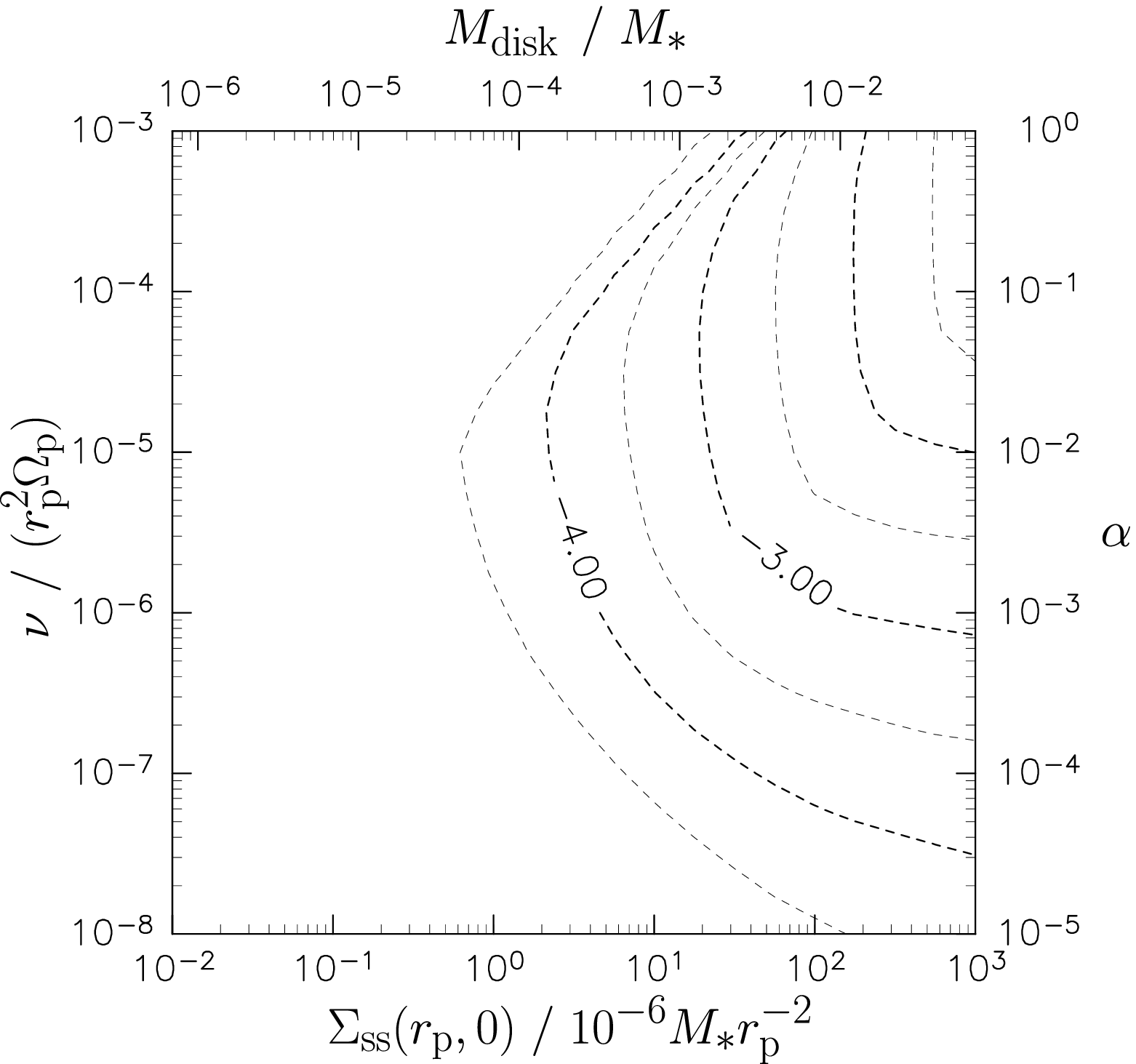}
\caption{Contour lines of $\log(M_{\rm final}/M_\ast)$ when global
viscous evolution is considered in the case with $\tau_{\rm dep} =
10^6$.  Horizontal axis is normalized surface density times $10^6$
(i.e., it corresponds to $\zeta$ when $\Sigma_{\infty,\rm init} =
\Sigma_{\rm ss}(r_{\rm p},0)$) and corresponding disk mass assuming
$R_{\rm out}=10r_{\rm p}$ is also shown on the top.  Vertical axis is
normalized viscosity and corresponding $\alpha$ is also shown on the
right.  Left panel shows the case with $h/r=0.1$ and right panel shows
when $h/r=0.032$.
\label{fig_final_mass_contour_with_ss}}
\end{figure}

% contour
Figure \ref{fig_final_mass_contour_with_ss} is a similar figure to
Fig.~\ref{fig_final_mass_contour} and shows the final mass of the
planet, $M_{\rm final}$, as a function of the initial surface density,
$\Sigma_{\infty, \rm init}$, and viscosity, $\nu$, in the case where
$h/r = 0.1$ (left panel) and $0.032$ (right panel), $R_{\rm out} = 10
r_p$, and $\tau_{\rm dep}=10^6 \Omega_{\rm p}^{-1}$.  The value of
$\tau_{\rm dep}$ corresponds to $\zeta = 10^6 (\Sigma_{\infty,\rm
init}/(M_\ast r_{\rm p}^{-2}))$, so that the ranges of this figure are
the same as those of Figure \ref{fig_final_mass_contour} for the
horizontal axis as well as the vertical axis.

Since we have adopted the self-similar solution for the surface density
evolution of the global disk, there appear two additional limits to the
final mass.
One arises from the appearance of the viscous timescale $\tau_{\rm
vis}$, in addition to the timescale of exponential decay $\tau_{\rm
dep}$.
When $\tau_{\rm vis} < \tau_{\rm dep}$, the effective disk lifetime is
$\tau_{\rm vis}$.  Thus, in order for planets to be massive, the growth
timescale $\tau_{\rm div}$ should be shorter than $\tau_{\rm vis}$, so
that the additional limit is given by $\tau_{\rm vis} = \tau_{\rm div}$:
\begin{equation}
\frac{\nu}{r_{\rm p}^2\Omega_{\rm p}}
\simeq
 7 \times 10^{-5}
 \kakkoi{M_{\rm p,init}}{10^{-5}M_\ast}{1/3}
 \kakkoi{h}{0.1r_{\rm p}}{-2}
 \kakkoi{\Sigma_\infty}{10^{-4}M_\ast / r_{\rm p}^2}{}
 \kakkoi{R_{\rm out}}{10r_{\rm p}}{},
\end{equation}
which runs from the lower left to the upper right in
Fig.\ref{fig_final_mass_contour_with_ss} because $\nu$ is linearly
proportional to the surface density.
As described above, the effective disk lifetime, $\tau_{\rm lifetime}$,
is the shorter one of the two timescales, so that we here define
\begin{equation}
\tau_{\rm lifetime}
= \left\{
   \begin{array}{ll}
    \tau_{\rm dep} & \mbox{if } \tau_{\rm dep} < \tau_{\rm vis} \\
    \tau_{\rm vis} & \mbox{if } \tau_{\rm dep} > \tau_{\rm vis}. \\
   \end{array}
\right.
\label{tau_lifetime}
\end{equation}

The other limit is the total mass of the
disk.  This limit does not appear in Fig.\ref{fig_final_mass_contour}
(i.e., the case without global viscous evolution).
This can be clearly seen in the case of $h/r = 0.032$ on the right panel
of Fig.~\ref{fig_final_mass_contour_with_ss} where contour lines are
vertical.
Note that the steep boundary (being equivalent to $\zeta = \zeta_{\rm
t}$) characterized by $\tau_{\rm div}=\tau_{\rm dep}$ that is seen in
Fig.\ref{fig_final_mass_contour} is not found in
Fig.~\ref{fig_final_mass_contour_with_ss}.
%
%\textcolor{blue}{, which is equivalent to $\zeta = \zeta_{\rm t}$ in \S
%\ref{sec_with_dissipation} (see eq.[\ref{zeta_t}])}.
%
This is simply because the two limits described above are more severe
than the condition of $\tau_{\rm div}=\tau_{\rm dep}$ in the cases shown
in Figure \ref{fig_final_mass_contour_with_ss}.
However, when $\tau_{\rm dep}$ is smaller, the boundary determined by
$\tau_{\rm div} = \tau_{\rm dep}$ moves rightward, so that the vertical
steep boundary determined by $\tau_{\rm div} = \tau_{\rm dep}$ emerges
even in the case with global disk evolution described in this
subsection.
%

%However $\tau_{\rm dep}$ is an independent variable
%({\it cf.} $\zeta$ described in \S \ref{sec_with_dissipation}), so that
%the boundary of $\tau_{\rm div}=\tau_{\rm dep}$ move rightward when
%$\tau_{\rm dep}$ is small.  In that case, the vertical steep boundary
%can be seen even in the global disk model case.

% Introduction of M_final vs r_p figure
\subsubsection{classification by semi-major axis}
\label{sec_classification}
\begin{figure}
\epsscale{0.9}
\plotone{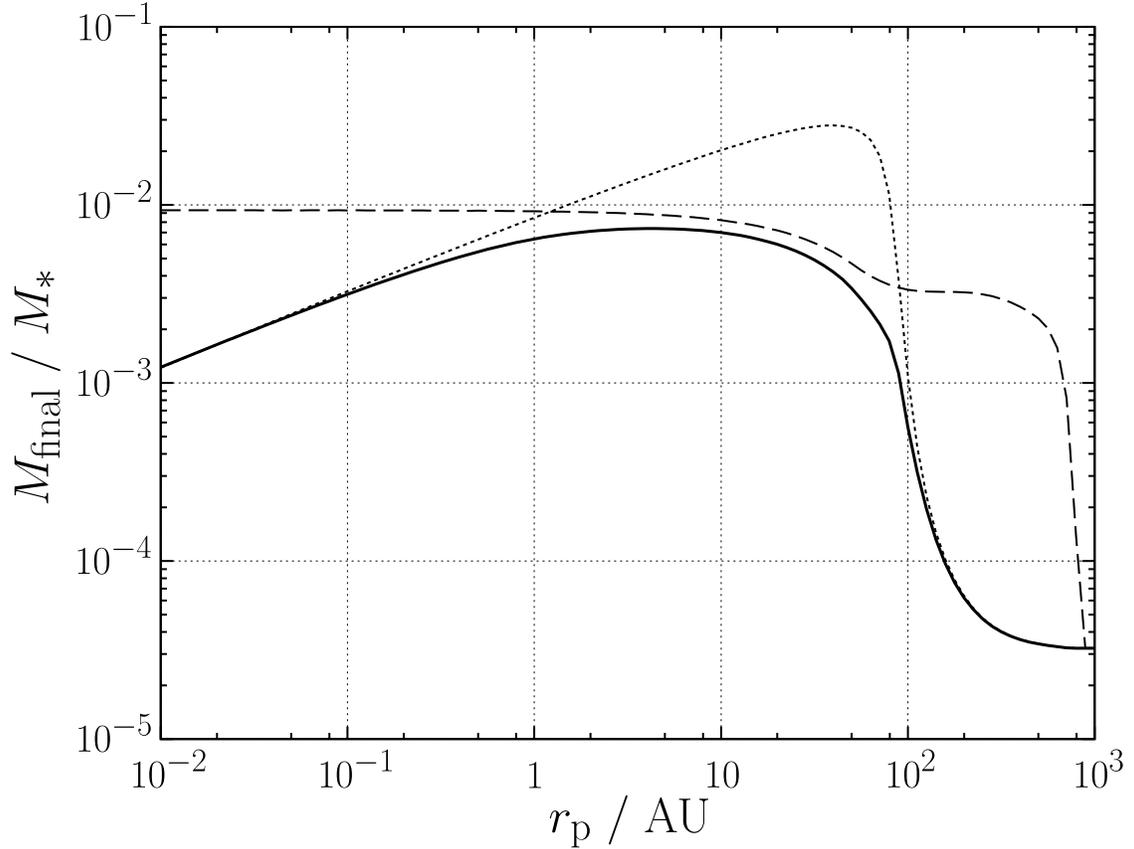}
\caption{Final mass of planets as a function of semi-major axis in AU.
Solid line shows standard case: $\alpha=0.01$, $h/r=0.032$ at 1AU,
$\tau_{\rm dep}=10^6$yr, $M_{\rm disk}=1.3\times 10^{-2}M_{\ast}$,
$R_{\rm out}=100$AU.  Dashed line shows $M_{\rm p,disk}$
(i.e., mass when all viscous-accreting gas is assumed to accrete onto
planets), and dotted line shows $M_{\rm p,ss}$ (i.e., mass when the
global viscous evolution with self-similar solution is not assumed).
Initial mass of the planets is set as $3.2 \times 10^{-5}M_\ast$, which
corresponds to 10 Earth masses in the solar mass system.
\label{fig_final_mass_vs_a-primitive}}
\end{figure}

In Figure~\ref{fig_final_mass_contour_with_ss}, the dependence of the
final mass was shown for the normalized disk parameters.  However, one
might want to know its dependence on semi-major axis.  We thus show the
final mass as a function of semi-major axis in
Figure~\ref{fig_final_mass_vs_a-primitive} for a typical case with
$\alpha=0.01$, $h/r=0.032$ at 1AU, $\tau_{\rm dep}=10^6$yr, $R_{\rm
out}=100$AU, $M_{\rm p,init} = 3.2 \times 10^{-5} M_\ast$, which
corresponds to 10 Earth masses in the solar mass system, and $M_{\rm
disk} = 0.013M_\ast$, which yields a surface density roughly 0.1 times
that of minimum mass solar nebula at 1AU\footnote{This surface density
is much lower than the minimum mass solar nebula model because the slope
of the surface density is shallower and the disk size is larger.}.  In
addition to the final mass given by equation~(\ref{M_final}), two kinds
of the mass defined by
\begin{equation}
 M_{\rm p,local} \equiv \int_0^\infty \dot{M}_{\rm p,local} dt
  \quad \mbox{and} \quad
 M_{\rm p,disk} \equiv \int_0^\infty \dot{M}_{\rm disk} dt
 \label{M_p_local-disk}
\end{equation}
are shown to clarify the limiting process for the final mass.

As shown in Figure~\ref{fig_final_mass_vs_a-primitive}, the final mass
(solid line) is determined basically by the smaller of $M_{\rm p,local}$
(dotted line) and $M_{\rm p,disk}$ (dashed line), although the limiting
process varies with time. Thus one finds three characteristic regions
that are divided at the intersection of the two curves.
%We can see from Figure \ref{fig_final_mass_vs_a-primitive} that the
%final mass \textcolor{blue}{(solid line)} is basically determined by the
%smaller one of $M_{\rm p,ss}$ \textcolor{blue}{(dotted line)} and
%$M_{\rm p,disk}$ \textcolor{blue}{(dashed line)}, \textcolor{red}{while
%both of $\dot{M}_{\rm p,ss}$ and $\dot{M}_{\rm disk}$ would be
%$\dot{M}_{\rm p}$ depending on the situation.}\footnote{ while
%$\dot{M}_{\rm p}$ is not determined only by either $\dot{M}_{\rm p,ss}$
%or $\dot{M}_{\rm disk}$ through the evolution.}  Thus the whole region
%can divide into three parts at the intersection of the two curves.

% Gap-limiting region
In the inner region ($r_p\lesssim 1$AU), $M_{\rm final}$ is almost equal
to $M_{\rm p,local}$ and is proportional to $r_p^{1/2}$.  The duration
of the pre-gap phase is shorter than the disk lifetime (i.e., $\tau_{\rm
div} < \tau_{\rm lifetime}$) in this region.  The gas accretion is thus
suppressed because of gap opening.  This region is hereafter called the
{\it gap-limiting region}.  In this situation the final mass can be
roughly estimated by integrating $\dot{M}_{\rm p}$ from $\tau_{\rm div}$
to $\tau_{\rm lifetime}$.  Since $\tau_{\rm div} < \tau_{\rm lifetime}$
in the gap-limiting region for the values of the parameters used in
drawing Fig.~\ref{fig_final_mass_vs_a-primitive}, $M_{\rm final}$ is
readily calculated as (see Appendix~\ref{Appendix-B})
\begin{eqnarray}
M_{\rm final}
& \simeq &
 \int_{\tau_{\rm div}}^{\tau_{\rm lifetime}}
  \frac{M_{\rm trans}}{t} dt
  \nonumber \\
& = &
 M_{\rm trans} \log \kakkoi{\tau_{\rm lifetime}}{\tau_{\rm div}}{}
\equiv
 M_{\rm final,gap}
% \equiv M_{\rm final,gap},
 \label{M_final_gap}
\end{eqnarray}
where $M_{\rm trans}$, defined by equation~(\ref{M_trans}), is a
planetary mass when $\Sigma_{\rm acc} = \Sigma_\infty/e$ and $2r_{\rm
H}\leq x_{\rm m}$.  Except for the weak dependence on the other
parameters included in the log term, the final mass given by
equation~(\ref{M_final_gap}) is determined principally by $\nu$ and
increases with $\nu$. This is because larger planetary mass is
required for the planet to open a gap in the case of higher viscosity.
Since $M_{\rm trans} \propto \nu/(r_p^2 \Omega_p)$ and $\nu \propto
r_p$ in our disk model, 
%as mentioned in \S~\ref{sec:model},
$M_{\rm trans} \propto \alpha r_p^{1/2}$, which is consistent with the
result shown in Fig.~\ref{fig_final_mass_vs_a-primitive}.
In the intermediate region (1AU$\lesssim r_p \lesssim$ 100AU),
Figure~\ref{fig_final_mass_vs_a-primitive} illustrates
\begin{equation}
M_{\rm final}
 \simeq M_{\rm p,disk}
\label{M_final_diff} % Caution!! The equation label has changed.
\end{equation}
In the region, viscous diffusion of the disk gas limits the gas
accretion onto the planet.  This region is hereafter called the
\textit{diffusion-limiting} region.
In this region, radial mass flux due to viscous diffusion is uniform
where $r_{\rm p} \ll R_{\rm out}$, so that the final mass is usually
insensitive to $r_{\rm p}$.
In the case with $\tau_{\rm vis} < \tau_{\rm dep}$, which is the case of
Fig.~\ref{fig_final_mass_vs_a-primitive}, most of the disk gas is
transferred toward the central star by viscous diffusion, thus
the final mass is roughly the disk mass ($\simeq 0.013 M_\ast$).
Hence we have
\begin{equation}
M_{\rm final,diff} \sim M_{\rm disk}.
\label{M_final_diff_1}
\end{equation}
On the other hand, when $\tau_{\rm vis} \gg \tau_{\rm dep}$, evaporation
dominates viscous diffusion in disk dissipation.  In this situation, the
final mass at $r \ll R_{\rm out}$ is given by
\begin{eqnarray}
M_{\rm final,diff}
& \simeq &
  \int_0^{\tau_{\rm dep}}
   |\dot{M}_{\rm disk}| dt  \nonumber \\
& \simeq &
  \frac{1}{2}
  \kakkoi{\tau_{\rm dep}}{\tau_{\rm vis}}{}
  M_{\rm disk}.
\label{M_final_diff_2}
\end{eqnarray}
Note that, even when $\tau_{\rm vis} \gg \tau_{\rm dep}$, the gas
accretion rate onto the planet is limited by viscous diffusion.

%In the intermediate region (1AU$\lesssim r \lesssim$ 100AU), where
%$M_{\rm p,disk} < M_{\rm p,ss}$, the disk tends to limit the mass flux
%to the planet before the planet open a gap deep enough to suppress the
%growth.  The mass flux of the disk is determined by viscous diffusion,
%thus we call this region {\it diffusion-limiting} region.  When
%$\tau_{\rm vis} \lesssim \tau_{\rm dep}$, whole disk dissipates by
%viscous evolution, so that
%\begin{equation}
%M_{\rm final,diff}
% \simeq M_{\rm p,disk}
% \sim M_{\rm disk}.
%\label{M_final_diff_1}
%\end{equation}
%The case shown in Figure \ref{fig_final_mass_vs_a-primitive} is
%$\tau_{\rm vis} < \tau_{\rm dep}$, thus the final mass is roughly the
%disk mass ($\simeq 0.013 M_\ast$).
%%%
%But when $\tau_{\rm vis} \gg \tau_{\rm dep}$, the disk starts to
%evaporate before the completion of viscous evolution.  In this case, the
%final mass at $r \ll R_{\rm out}$ is
%\begin{eqnarray}
%M_{\rm final,diff}
%& \simeq &
%  \int_0^{\tau_{\rm dep}}
%   |\dot{M}_{\rm disk}| dt  \nonumber \\
%& \simeq &
%  \frac{1}{2}
%  \kakkoi{\tau_{\rm dep}}{\tau_{\rm vis}}{}
%  M_{\rm disk}
%\label{M_final_diff_2}
%\end{eqnarray}

%
% time limited region
In the outer region ($r_p \gtrsim$ 100AU), the final mass suddenly
decreases to $M_{\rm p,init}$.  This is because the planetary growth
finishes in the pre-gap phase because of the dissipation of the disk
gas (i.e., $\tau_{\rm div} \gtrsim \tau_{\rm lifetime}$).
We call this region the {\it no-growth region}.
%At outer region ($r\gtrsim$ 100AU), the final mass suddenly decrease
%with radius.  This is because the pre-gap-formed phase growth does not
%finish within the lifetime of the disk (i.e., $\tau_{\rm div} \gtrsim
%\tau_{\rm lifetime}$).  \textcolor{blue}{We call this region {\it
%no-growth} region.}

%%%
% Boundaries between the three regions
The positions of the boundaries between the three regions are given as
follows.
% r_b
At the boundary between the gap-limiting region and the
diffusion-limiting region (its position being denoted by $r_{\rm b}$),
$M_{\rm final,gap}=M_{\rm final,diff}$.  When $\tau_{\rm vis} <
\tau_{\rm dep}$, from equations~(\ref{M_final_gap}) and
(\ref{M_final_diff_1}), we obtain
\begin{equation}
r_{\rm b}
 \simeq
  5
  \kakkoi{\alpha}{10^{-2}}{-2}
  \kakkoi{h_{\rm 1AU}}{10^{-1.5}\rm AU}{-4}
  \kakkoi{M_{\rm disk}}{10^{-2}M_\ast}{2}
  \kakkoi{\log(\tau_{\rm vis}/\tau_{\rm div})}{5}{-2}.
\label{r_b_1}
\end{equation}
Here we have assumed the log factor in equation~(\ref{M_final_gap}) is
constant.
When $\tau_{\rm vis} > \tau_{\rm dep}$, from
equations~(\ref{M_final_gap}) and (\ref{M_final_diff_2}), we have
\begin{equation}
r_{\rm b}
 \simeq
  1
  \kakkoi{\alpha}{10^{-2}}{-2}
  \kakkoi{h_{\rm 1AU}}{10^{-1.5}r_{\rm 1AU}}{-4}
  \kakkoi{M_{\rm disk}}{10^{-2}M_\ast}{2}
  \kakkoi{\tau_{\rm dep}}{\tau_{\rm vis}}{2}
  \kakkoi{\log(\tau_{\rm dep}/\tau_{\rm div})}{5}{-2}.
\label{r_b_2}
\end{equation}
%We then consider the boundary positions of the three regions.
%%% r_b
%The boundary between the gap-limiting region and diffusion-limiting
%region is determined by $M_{\rm final,gap} = M_{\rm final,diff}$ and the
%position is denoted by $r_{\rm b}$.  For $M_{\rm final,gap}$, the factor
%of log in eq.[\ref{M_final_gap}] is assumed to be constant because of
%the weak dependence.  When $\tau_{\rm vis} < \tau_{\rm dep}$, $M_{\rm
%final,diff}$ is given by eq.[\ref{M_final_diff_1}], thus we obtain
%\begin{equation}
%r_{\rm b}
% \simeq
%  5
%  \kakkoi{\alpha}{10^{-2}}{-2}
%  \kakkoi{h_{\rm 1AU}}{10^{-1.5}\rm AU}{-4}
%  \kakkoi{M_{\rm disk}}{10^{-2}M_\ast}{2}.
%\label{r_b_1}
%\end{equation}
%When $\tau_{\rm vis} > \tau_{\rm dep}$, we have
%\begin{equation}
%r_{\rm b}
% \simeq
%  1
%  \kakkoi{\alpha}{10^{-2}}{-2}
%  \kakkoi{h_{\rm 1AU}}{10^{-1.5}r_{\rm 1AU}}{-4}
%  \kakkoi{M_{\rm disk}}{10^{-2}M_\ast}{2}
%  \kakkoi{\tau_{\rm dep}}{\tau_{\rm vis}}{2}.
%\label{r_b_2}
%\end{equation}
%
% r_edge
The location of the boundary between the diffusion-limiting region and
the no-growth region is basically determined by $\tau_{\rm div} \sim
\tau_{\rm lifetime}$ and is denoted by $r_{\rm e}$.  When $\tau_{\rm
vis} < \tau_{\rm dep}$, from $\tau_{\rm div}=\tau_{\rm vis}$, we obtain
\begin{equation}
r_{\rm e}
 \simeq
  70
  \kakkoi{M_{\rm p,init}}{10^{-5}M_\ast}{1/3}
  \kakkoi{\Sigma_{\infty,\rm 1AU}}{10^{-5}M_\ast / \rm 1AU^2}{}
  \kakkoi{h_{\rm 1AU}}{10^{-1.5}\rm AU}{-4}
  \kakkoi{\alpha}{10^{-2}}{-1}
  \kakkoi{R_{\rm out}}{\rm 100AU}{}
  {\rm AU}.
\label{r_edge_1}
\end{equation}
When $\tau_{\rm vis} > \tau_{\rm dep}$, from $\tau_{\rm div}=\tau_{\rm
dep}$, we obtain
\begin{equation}
r_{\rm e}
 \simeq
  130
  \kakkoi{M_{\rm p,init}}{10^{-5}M_\ast}{1/3}
  \kakkoi{\Sigma_{\infty,\rm 1AU}}{10^{-5}M_\ast / \rm 1AU^2}{}
  \kakkoi{h_{\rm 1AU}}{10^{-1.5}\rm AU}{-2}
  \kakkoi{\tau_{\rm dep}}{10^6 \rm yr}{}
  {\rm AU}.
\label{r_edge_2}
\end{equation}

%%% r_edge
%The boundary between the diffusion-limiting region and no-growth region
%is basically determined where $\tau_{\rm div} \sim \tau_{\rm lifetime}$ and
%we denote the position as $r_{\rm edge}$.  When $\tau_{\rm vis} <
%\tau_{\rm dep}$,
%\begin{equation}
%r_{\rm edge}
% \simeq
%  70
%  \kakkoi{M_{\rm p,init}}{10^{-5}M_\ast}{1/3}
%  \kakkoi{\Sigma_{\infty,\rm 1AU}}{10^{-5}M_\ast / \rm 1AU^2}{}
%  \kakkoi{h_{\rm 1AU}}{10^{-1.5}\rm AU}{-4}
%  \kakkoi{\alpha}{10^{-2}}{-1}
%  \kakkoi{R_{\rm out}}{\rm 100AU}{}
%  {\rm AU}.
%\label{r_edge_1}
%\end{equation}
%When $\tau_{\rm vis} > \tau_{\rm dep}$,
%\begin{equation}
%r_{\rm edge}
% \simeq
%  130
%  \kakkoi{M_{\rm p,init}}{10^{-5}M_\ast}{1/3}
%  \kakkoi{\Sigma_{\infty,\rm 1AU}}{10^{-5}M_\ast / \rm 1AU^2}{}
%  \kakkoi{h_{\rm 1AU}}{10^{-1.5}\rm AU}{-2}
%  \kakkoi{\tau_{\rm dep}}{10^6 \rm yr}{}
%  {\rm AU}.
%\label{r_edge_2}
%\end{equation}

\begin{figure}
\epsscale{0.9}
\plotone{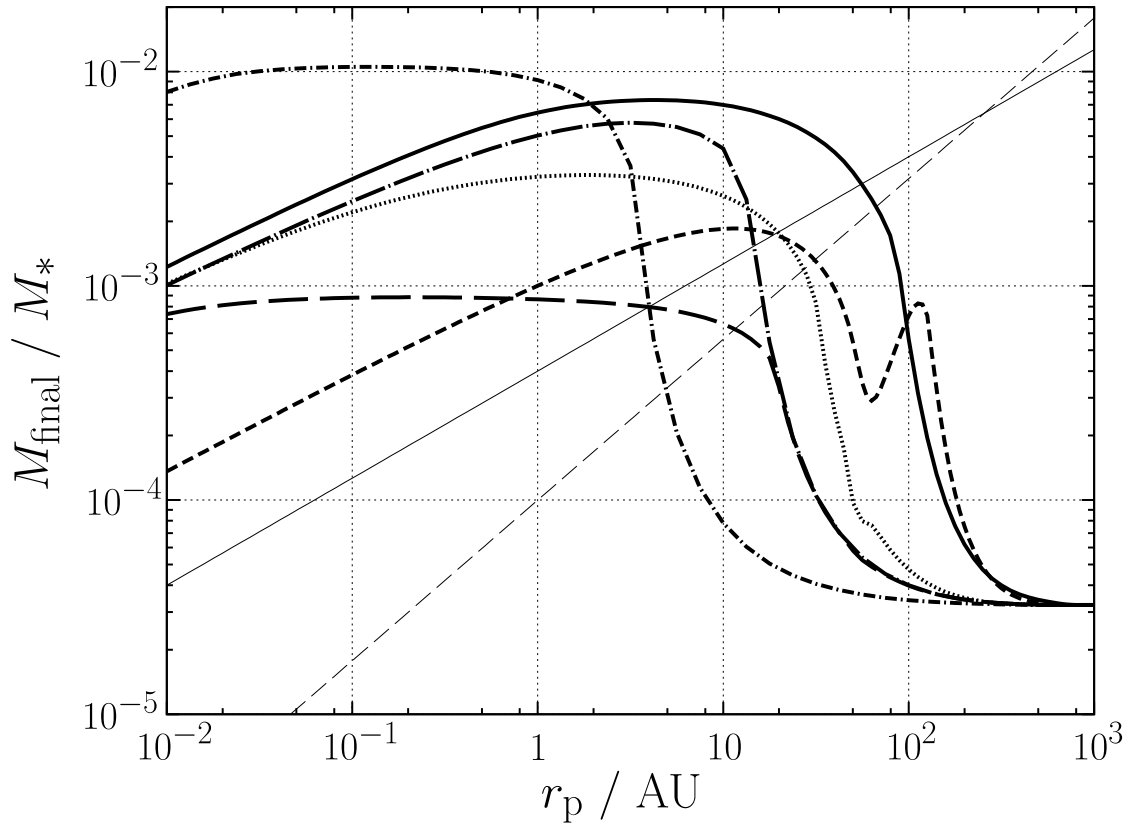}
\caption{Final mass of planets as a function of semi-major axis in AU
for several disk models.  Thick solid line shows standard case as
described in Figure \ref{fig_final_mass_vs_a-primitive}.  Thick
long-dashed line shows $M_{\rm disk}$ is 10 times smaller case, thick
short-dashed line shows viscous $\alpha$ is 10 times smaller than the
standard case, thick dotted line shows $\tau_{\rm dep}$ is 10 times
shorter case, thick dot-short-dashed line shows the case when scale
height is 3 times higher, and thick dot-long-dashed line shows the case
when gas accretion rate is reduced by a factor of 10.  Thin solid line
and thin dashed line show the masses determined by the viscous and
thermals condition ($\alpha=0.01$), respectively.
\label{fig_final_mass_vs_a}}
\end{figure}

%%%
Figure \ref{fig_final_mass_vs_a} shows the final mass for several
different values of the disk parameters.  The discussion above suffices
to understand this figure.
%
%Next we briefly describe the dependence of the parameters on the results.
%Figure \ref{fig_final_mass_vs_a} shows the final mass as a function of
%semi-major axis with different disk parameters.
%%%
%There are several lines corresponding to the cases with different values
%for the parameters.  We can see that the final mass can change greatly
%depending on the parameters.
%%%
%We have already showed what determined the final mass and the boundary
%depending on the regions.
%%%
%In addition, this figure is only some two-dimensional cross sections in
%a six-dimensional (five parameters plus the final mass) phase space,
%thus it is practically impossible to describe all the dependence.
%%%
%Instead, we here comment on two points.
%
%%% r_edge
%One of the important dependences is what determines $r_{\rm edge}$.  The
%position $r_{\rm edge}$ depends on several parameters as seen in
%eqs.[\ref{r_edge_1}] and [\ref{r_edge_2}], depending on the relationship
%between $\tau_{\rm vis}$ and $\tau_{\rm dep}$.
%%%
%The ranges of $\Sigma_{\infty,\rm 1AU}$ and $\alpha$ are likely to
%change more than a factor of 10, which is directly reflected in $r_{\rm
%edge}$.
%%%
%In addition, $h_{\rm 1AU}$ is a most sensitive parameter on $r_{\rm
%edge}$ even though the possible range of $h_{\rm 1AU}$ is not very wide.
%When $\tau_{\rm vis} < \tau_{\rm dep}$, $r_{\rm edge}$ is inversely
%proportional to the fourth power of $h_{\rm 1AU}$.  Out of four, two is
%arisen from $\dot{A}$ through $\tau_{\rm div}$ and the other two is from
%$\tau_{\rm vis}$, which is proportional to $h^{-2}$ in the $\alpha$
%
% Comment on ``Hekomi''
Note that there is a hollow at $\sim$ 50~AU in the case where the value
of $\alpha$ is lower by a factor of 10 than its nominal value (thick
short-dashed line).  This is, however, artificial.  In the self-similar
solution we have adopted, the direction of the radial gas flow due to
viscous diffusion changes at $r=(R_{\rm out}/2) \tilde{\tau}_{\rm ss}$,
which means that the mass flux is zero at the point, and the accretion
rate onto the planet is also zero accordingly (see
eq.~[\ref{Mdot_p_min}]).
Since $\tilde{\tau}_{\rm ss}$ increases with time (see
eq.[\ref{tilde_tau_ss}]), the point moves outward with time.  When
$\tau_{\rm dep} > \tau_{\rm vis}$, the point moves significantly with
time and integration of mass flux at a particular point eliminates the
effect.  In contrast, when $\tau_{\rm dep} \ll \tau_{\rm vis}$, the
point hardly changes before the exponential depletion; the planet around
$r=R_{\rm out}/2$ thus cannot grow significantly.

Another important aspect is on the final mass in the gap-limiting
region.  We can see that the $r_{\rm p}$ dependence of $M_{\rm final}$
agrees with that of the mass given by the viscous condition for gap
opening \citep{LP79,LP93}:
\begin{eqnarray}
M_{\rm p,vis}
&=& 40
    \kakkoi{\nu}{r^2\Omega}{}
    M_\ast \label{M_p_vis} \\
&=& 4\times 10^{-4}
    \kakkoi{\alpha}{10^{-2}}{}
    \kakkoi{r}{\rm 1AU}{1/2}
    M_\ast.
\end{eqnarray}
The dependence also agrees with that of $M_{\rm trans}$, which is the
mass when $\Sigma_{\rm acc}$ is $(1/e)\Sigma_\infty$ in the case of
$2r_{\rm H} \geq x_{\rm m}$ (see eq.[\ref{M_trans}]), and re-written as
\begin{equation}
M_{\rm trans}
\simeq
 9 \times 10^{-4}
 \kakkoi{\alpha}{10^{-2}}{}
 \kakkoi{r}{\rm 1AU}{1/2}
 M_\ast.
\end{equation}
Both of the two masses indicate gap-forming masses and they agree with
each other within a factor of about 2.  However the final mass shown in
our model is larger by a factor of 10 than $M_{\rm p,vis}$ or $M_{\rm
trans}$.  This is because the two masses correspond to the masses at
which a gap is about to form and does not mean the masses at which
growth stops.  A further increase in $M_{\rm p}$ after reaching $M_{\rm
p,vis}$ or $M_{\rm trans}$ is reflected in the $\log$ term in equation
(\ref{M_final_gap}).

%======================================================================
\subsection{Remarks on the uncertainties in the model}\label{sec_remarks}
%
% Assumption of equilibrium surface density
We have assumed that the distribution of the disk gas around the
planetary orbit (i.e., the surface density) is always in the equilibrium
state that is determined by the balance between viscous stress and
gravitational scattering by the planet.
However, for the equilibrium state to be achieved, non-equilibrium
distributions must be relaxed by diffusion. The diffusion timescale,
$\tau_{\rm vis,local}$, is estimated by the gap width ($\sim 2\times
2r_{\rm H}$) divided by the viscosity as
\begin{equation}
\tau_{\rm vis,local}
 = 7.7 \times 10^3
   \kakkoi{M_{\rm p}}{10^{-3}M_\ast}{2/3}
   \kakkoi{\nu}{10^{-5}r_{\rm p}^2\Omega_{\rm p}}{-1}
   \Omega_{\rm p}^{-1}.
\end{equation}
This timescale should be compared with the typical growth timescale
(i.e., $\tau_{\rm div}$; see eq.~[\ref{tau_div}]).  The comparison
indicates that $\tau_{\rm vis,local} < \tau_{\rm div}$ in most of the
cases presented in this paper. Hence the assumption is appropriate
unless quite low viscosities (yielding large $\tau_{\rm vis,local}$) or
high surface densities (yielding small $\tau_{\rm div}$) are considered.

% Uncertainty arisen from the formula of TW02's
The empirical formula for the accretion rate based on local
two-dimensional isothermal hydrodynamic simulations (TW02) can be
different from that derived based on more realistic simulations
including, for example, three-dimensional accretion flow
\citep{DKH03,Bate03}, non-isothermal equation of state (TW02), and
magnetic field \citep{Machida06}.
The modification would change the form of $\tau_{\rm
div}$, which would yield quantitatively different results.  However,
even in that case, following our prescription given in this paper, one
can easily calculate the mass evolution of planets in a similar way.
The formula given by TW02 can be considered as the highest limit; thus,
for example, $r_{\rm e}$ is expected to shift inward if more realistic
models for the accretion rate are used.
We compare the accretion rate eq.[\ref{mdot}] with those obtained by
global simulations \citep{Kley99,LD06}.  Although the value of our
accretion rate are usually larger (by up to a factor of 10) than those
given by the simulations, the dependence on viscosity and planetary mass
is consistent.
%The difference may arise from the too short calculation
%time for their hydrodynamic simulations.

% Geometrical limit on the accretion rate
We have not taken into account a limit in terms of geometry of the
accretion flow, namely a within which equation (\ref{Adot}) is
applicable.  We can make a simple estimate of the maximum accretion rate
if we assume all the gas within $0 \leq x \leq 2\sqrt{3}r_{\rm H}$ in
$y>0$ and $-2\sqrt{3}r_{\rm H} \leq x \leq 0$ in $y<0$ accrete to the
planet, where $x=2\sqrt{3}r_{\rm H}$ at $y \gg r_{\rm H}$ is the point
where potential energy on the rotating frame is the same as that at the
Lagrange points L1 or L2.  In this case, the maximum accretion rate
(normalized by surface density) is given by $3^{1/3}6 r_{\rm p}^2
\Omega_{\rm p} (M_{\rm p}/M_\ast)^{2/3}$ and thus $\dot{A}$
(eq.[\ref{Adot}]) becomes larger than the maximum value when
\begin{equation}
M_{\rm p}
 > 5.6 \times 10^{-3} M_\ast
   \kakkoi{h}{0.032r_{\rm p}}{3}
 \equiv M_{\rm p,crit}.
 \label{M_p_crit}
\end{equation}
Although planetary mass can be larger than $M_{\rm p,crit}$ depending on
the parameters, such a massive planet should already has a deep gap, so
that the accretion rate is reduced greatly when the planetary mass
reaches $M_{\rm p,crit}$.  Consequently the geometric effect has no
significant influence on the final mass.
% (added in the second revised version)
Note that the discussion here is based on an approximation that
streamlines can be well described by particle motions on the framework
of the restricted three-body problem.  Since this approximation is not
valid when $r_{\rm H}<h$ \citep{Masset06}, it is not applicable for
small mass planet's cases.  However, we consider the upper limit of
applicable mass, where the planet mass is, in most cases, large enough
to satisfy $r_{\rm H}>h$, so that the approximation is valid for the
situations that we consider here.

% Type II migration
In this paper, we have put off the issue of planetary migration.  Since
we consider a phase in which a gap exists around the planet, we may have
to consider the type II migration \citep[e.g.][]{Ward97,Ward00},
especially in the gap-limiting region (e.g., $r_p<1$~AU in Figure
\ref{fig_final_mass_vs_a-primitive}).
The deep gap created by the planet blocks the accretion flow toward the
central star.  It follows that the planet is pushed inward by the gas
exterior to the planet's orbit.  Inclusion of the effect of planetary
migration is our future work.

%In this paper, we have neglected the effects of planetary migration.  We
%modeled the phase after the onset of gas capture, so that type II
%migration may be considered and gap-limiting region (e.g., $r<1$AU in
%Figure \ref{fig_final_mass_vs_a-primitive}) would be the case.
%%%
%In the case with viscous evolution of the disk described in \S
%\ref{sec_self-similar}, the gas exterior to the planetary orbit would
%flow through the planet if the planet were not there, and the total mass
%of disk gas passing through the planetary orbit would be larger than the
%final mass shown in the figure.  But, on the other hand, the planet
%makes a deep gap that blocks the gas accretion through the planetary
%orbit toward the central star.  Thus it is natural to consider that the
%planet in the situation is pushed inward by the gas exterior to the
%planet's orbit.  Inclusion of planetary migration is our future work.

%

%%%%%%%%%%%%%%%%%%%%%%%%%%%%%%%%%%%%%%%%%%%%%%%%%%%%%%%%%%%%%%%%%%%%%%%
\section{Summary}\label{sec_summary}

%%%
To gain a systematic understanding of the final masses of gas giant
planets, we have simulated the long-term accretion of gas giant planets
after the onset of the supercritical gas accretion in a variety of
situations, depending on four disk parameters such as disk mass,
viscosity, scale height, and semi-major axis.  To do so, we have made a
semi-analytical model to simulate the mass evolution, which enabled us
to study the final mass of gas giant planets for extensive ranges of all
the parameters.

%%% TAKA'S
%We have calculated the mass evolution of gas giant planets embedded in
%gas disks using semi-analytical way to gain a systematic understanding
%on the final mass of gas giant planets under wide varieties of
%situation, depending on various parameters such as disk mass, viscosity,
%scale height, and semi-major axis.
%
%We made a model to treat the mass evolution using semi-analytical way,
%which enables us to study dependence of extensive range of all the
%parameters.
%
%We considered the phase after the onset of gas accretion onto the solid
%core of protoplanets.  The migration of the planets was neglected.

%%%
We have first made a 1D analytical model of the equilibrium surface
density profile around a protoplanet, from consideration of the balance
of torque and the dynamical stability (\S \ref{sec_surface_density}).
Combining the surface density profile with an empirical formula for gas
accretion rate that was obtained on the basis of hydrodynamic
simulations by TW02, we have obtained a formula for gas
accretion rate as a function of planetary mass, viscosity, scale height,
and unperturbed surface density (\S \ref{sec_accretion_rate}).  We have
then integrated the gas accretion rate numerically with respect to time
to simulate the long-term accretion of gas giant planets (\S 4). To
understand the basic behavior of the planetary accretion, we have
explored two simple cases with no disk dissipation (\S
\ref{sec_without_dissipation}) and with exponentially-decreasing surface
density (\S \ref{sec_with_dissipation}). Finally, we have simulated the
long-term accretion of gas giant planets embedded in a viscously
evolving and evaporating disk to obtain the final mass of gas giant
planets as a function of semi-major axis (\S \ref{sec_self-similar}).
We have consequently found the following three different regions
depending on limiting processes on the final mass.

%%% TAKA'S
%We first obtained equilibrium surface density profile with
%one-dimensional approximation considering balance between planetary
%gravity and viscosity or gas pressure.
%
%Using a normalized formula of the gas accretion rate onto planets based
%on hydrodynamic simulations (TW02) in addition to the obtained
%surface density profile, we constructed an formula to describe
%normalized gas accretion rate as a function of planetary mass,
%viscosity, scale height, and unperturbed surface density.
%%%
%We then integrated the accretion rate numerically with respect to time
%to understand typical long-term mass evolutions of planets in
%no-depletion disks.
%
%After that, exponential disk depletion, which mimicked photo-evaporation
%process for example, was added in the model, enabling us to determine
%the final mass of the planets as a function of the parameters normalized
%locally at the planetary position.
%
%%%
%Based on the normalized calculation, we next consider global evolving
%disk for $\alpha$ viscous model to obtain the final mass as a function
%of semi-major axis of the planet.
%
%We found three regions depending on limiting processes on the final mass.

%
In the inner region ($r_{\rm p} \lesssim r_{\rm b}$; see
eqs.~[\ref{r_b_1}] \& [\ref{r_b_2}]), the planet grows quickly to form a
deep gap to suppress the gas accretion from the disk by itself within
the disk lifetime (\textit{gap-limiting} region).  We have found the
final mass in this region is roughly 10 times larger than that
determined by the viscous condition for gap opening \citep{LP93}.
%or $M_{\rm trans}$ (see eq.~[\ref{M_trans}] for its definition).  
This is because the condition for gap opening only expresses the
condition when a gap begins to form, and is by no means equivalent to
the condition that the growth is terminated.
%

%%% TAKA'S
%Most inner region ($r \lesssim r_{\rm b} \sim$1AU) is gap-limiting
%region where the planets can grow quickly to form a deep gap to suppress
%the gas accretion from the disk by themselves within the disk lifetime.
%The final mass of this region is roughly 10 times larger than that
%determined by viscous condition for gap opening \citep{LP93} or $M_{\rm
%trans}$.  This is because the condition for gap opening expresses the
%condition when a gap begins to form, and does not necessary mean when
%the growth is terminated.

%
In the intermediate region ($r_{\rm b} \lesssim r_{\rm p} \lesssim
r_{\rm e}$; see eqs.~[\ref{r_edge_1}] \& [\ref{r_edge_2}]), radial
transfer of the disk gas toward the planetary orbit limits the gas
accretion before the planet opens a deep gap; the final mass is thus
limited by viscous diffusion of the disk
(\textit{diffusion-limiting} region).
We have found that when the evaporation timescale $\tau_{\rm dep}$ is
shorter than the viscous-diffusion timescale $\tau_{\rm vis}$, the
relationship between the final mass $M_{\rm final}$ and the disk mass
$M_{\rm disk}$ is given by $M_{\rm final} \sim (1/2)(\tau_{\rm
dep}/\tau_{\rm vis})M_{\rm disk}$, whereas $M_{\rm final} \sim M_{\rm
disk}$ when $\tau_{\rm dep} > \tau_{\rm vis}$.
%

%%%TAKA'S
%Intermediate region ($r_{\rm b} \lesssim r \lesssim r_{\rm edge} \sim$
%100AU), radial gas transfer toward the planetary orbit in the disk
%starts to limit the gas accretion before the planet open a deep gap,
%thus the final mass is limited by viscous diffusion of the disk
%(diffusion-limiting region).
%
%When the timescale of exponential depletion $\tau_{\rm dep}$
%(corresponding to photoevaporation, for example) is shorter than that of
%viscous evolution $\tau_{\rm vis}$, the final mass becomes $\sim
%(1/2)(\tau_{\rm dep}/\tau_{\rm vis})M_{\rm disk}$, whereas the final
%mass is $\sim M_{\rm disk}$ when $\tau_{\rm dep} > \tau_{\rm vis}$.

%
In the outer region ($r \gtrsim r_{\rm e}$), the planet captures only a
tiny amount of gas by the time the disk gas completely dissipates
(\textit{no-growth} region). Saturn and possibly Uranus/Neptune are
likely to have experienced the situation.
%

%TAKA'S 
%In outer region ($r \gtrsim r_{\rm edge}$), the growth timescale
%$\tau_{\rm div}$ is shorter than the disk lifetime, so that the planet
%cannot capture significant amount of gas.

%
In this study, we have gained a clear understanding of the final masses
of gas giant planets, deriving analytical expressions for them in three
characteristic regions
(eqs.~[\ref{M_final_gap}]--[\ref{M_final_diff_2}]) and the locations of
the boundaries between the three regions
(eqs.[\ref{r_b_1}]--[\ref{r_edge_2}]).  To understand the mass-period
distribution of gas giant planets in extrasolar systems found by
radial-velocimetry, we need to take several additional processes into
consideration.  Especially, planets in the gap-limiting region would be
susceptible to the type II migration, because the gas exterior to the
planetary orbit blocked by the gap pushes the planet inward.  Inclusion
of planetary migration is our future work.  Also, inclusion of core
accretion processes and the gas accretion process governed by the
Kelvin-Helmholtz contraction of the envelope is needed especially to
determine the initial mass and the origin of time of our model.

%%%TAKA'S
%Note that the boundary position between the three regions depends on the
%parameters (see eqs.[\ref{r_b_1}]--[\ref{r_edge_2}]).  These positions
%are important to characterize the resultant configurations of planetary
%systems, but the parameters are poorly understood, thus it is difficult
%to give a reliable value within a narrow range.
%%
%We should also note that the planets in gap-limiting region would be
%susceptible to type II migration because gas outside of the planetary
%orbit is blocked by the gap even though the disk gas tries to accrete
%inward.  Inclusion of the planetary migration is our future works.

%%% Applications

%
Although the final masses of gas giant planets were focused on in this
paper, the accretion process for reaching the final mass is also
important to resolve issues relevant to planet formation. Growing
giant planets dynamically affect other bodies in a planetary
system. Satellites are likely to form in sub-disks around accreting
gas giant planets \citep[e.g.][]{CW02,CW06}.
The long-term accretion of gas giant planets may affect the internal
structure and evolution of isolated young gas giants \citep{Marley07}.
This would be important for future direct detection of young gas
giants.

\acknowledgments

We are grateful to S. Ida for fruitful discussion and continuous
encouragement. We also thank H. Tanaka for critical comments on our
modeling.  Valuable comments and suggestions from the anonymous referee
were quite helpful in improving this paper.  This work was supported by
Ministry of Education, Culture, Sports, Science and Technology of Japan
(MEXT), Grand-in-Aid for Scientific Research on Priority Areas,
``Development of Extra-solar Planetary Science'' (MEXT-16077202).


%% in parentheses, after the keyword, but they will not be verified.

%{\it Facilities:} \facility{Nickel}, \facility{HST (STIS)}, \facility{CXO (ASIS)}.



\appendix

%%%%%%%%%%%%%%%%%%%%%%%%%%%%%%%%%%%%%%%%%%%%%%%%%%%%%%%%%%%%%%%%%%%%%%%
\section{Disk properties}
% Convert the normalized quantities in real dimension.
Typical values of the disk parameters in our model are summarized with
their dependence on the semi-major axis of the planet.  We assume that the
disk is optically thin and in the radiative equilibrium with the stellar
radiation.  The disk temperature is given by \citep{Hay81}
\begin{equation}
T
 = 280{\rm K}
   \kakkoi{r_{\rm p}}{\rm 1AU}{-1/2}
   \kakkoi{L_\ast}{L_\odot}{1/4},
\label{Hayashi model}
\end{equation}
where $L_\ast$ and $L_\odot$ is luminosities of the central star and
the sun, respectively.  Hence the normalized scale height (i.e., the
aspect ratio of the disk) is
\begin{equation}
\frac{h}{r_{\rm p}}
 = 0.033
   \kakkoi{r_{\rm p}}{\rm 1AU}{1/4}.
 \label{aspect_ratio}
\end{equation}
Note that the ratio is independent of the stellar mass because we assume
the stellar luminosity is proportional to the fourth power of stellar
mass, which is observationally known in the case of main sequence stars.

On the $\alpha$-prescription ($\alpha \equiv \nu/(ch)$), which is widely
used to describe poorly-known turbulent viscosity in the disks
\citep{SS73}, the normalized viscosity coefficient with the temperature
distribution given by equation~(\ref{Hayashi model}) is
\begin{equation}
\frac{\nu}{r_{\rm p}^2 \Omega_{\rm p}}
 = 1.1 \times 10^{-5}
   \kakkoi{\alpha}{10^{-2}}{}
   \kakkoi{r_{\rm p}}{\rm 1AU}{1/2},
 \label{viscosity}
\end{equation}
If we assume that disk surface density is proportional to the mass of
the central star, normalized surface density is
\begin{equation}
\frac{\Sigma}{M_\ast/r_{\rm p}^2}
 = 1.9 \times 10^{-4}
   \kakkoi{r_{\rm p}}{\rm 1AU}{1/2}
   \kakkoi{f}{1}{},
\end{equation}
where $f$ is a factor relative to the surface density of the
minimum mass solar nebula model when the mass of the central star
equals to that of the Sun.
The dissipation timescale of the disk is
\begin{equation}
\frac{\tau_{\rm dep}}{\Omega_{\rm p}^{-1}}
 = 2\pi \times 10^6
   \kakkoi{\tau_{\rm dep}}{10^6 \rm yr}{}
   \kakkoi{r_{\rm p}}{\rm 1AU}{-3/2}
   \kakkoi{M_\ast}{M_\odot}{-1/2},
\end{equation}
where $M_\odot$ is the stellar mass.
The accretion rate is
\begin{equation}
\frac{\dot{M}_{\rm p}}{M_\ast \Omega_{\rm p}}
 = 4.8 \times 10^{-9}
   \kakkoi{\dot{M}_{\rm p}}{10^{-2}M_\oplus / {\rm yr}}{}
   \kakkoi{M_\ast}{M_\odot}{-3/2}
   \kakkoi{r_{\rm p}}{\rm 1AU}{3/2}
\end{equation}

%%%%%%%%%%%%%%%%%%%%%%%%%%%%%%%%%%%%%%%%%%%%%%%%%%%%%%%%%%%%%%%%%%%%%%%
\section{Approximate solutions for the accretion rate in the post-gap phase }
\label{Appendix-B}
%\section{Accretion rate as a explicit function of time after gap formation}
%\label{Appendix-B}
The surface density of the gas in the accretion band at $x \sim 2r_{\rm
H}$, $\Sigma_{\rm acc}$, is given by $\Sigma_{\rm vis}$, when the
accretion band exists in the region with the dynamically stable profile
of surface density determined by the balance between viscous diffusion
and scattering due to the planet, namely, when $2r_{\rm H} \geq x_{\rm
m}$.
Substituting $x=2r_{\rm H}$ into equation (\ref{Sigma_vis}), one obtains
\begin{equation}
\Sigma_{\rm acc}
 = \Sigma_{\rm vis}(2r_{\rm H})
 = \Sigma_\infty
   \exp\left(
    -\frac{M_{\rm p}}{M_{\rm trans}}
   \right),
\label{sigma_acc_in_post_gap_phase_vis}
\end{equation}
where
\begin{equation}
M_{\rm trans}
 = 27\pi
   \kakkoi{\nu}{r_{\rm p}^2 \Omega_{\rm p}}{}
   M_\ast.
 \label{M_trans}
\end{equation}
Using equation~(\ref{sigma_acc_in_post_gap_phase_vis}), one can rewrite
$\dot{M}_p = \Sigma_{\rm acc} \dot{A}$ as
\begin{equation}
\dot{M}_{\rm p}
 \simeq \frac{M_{\rm trans}}{t}
 \label{mdot_analytic_1}
\end{equation}
for large $t$ and $M_{\rm p}$, if one neglects the weak dependence of
$\dot{A}$ on $M_p$.
%is  %equation (\ref{mdot}) is actually 
%the product of $\Sigma_{\rm acc}$ and $\dot{A}$
%\textcolor{green}{(see eq.~[\ref{mdot}])}. 
%the dependence of $M_{\rm p}$ on
%$\dot{A}$ is much weaker than that on $\Sigma_{\rm acc}$ in the
%post-gap-formed phase.  We hence treat the accretion rate as
%\begin{equation}
%\dot{M}_{\rm p}
% = C \exp\left( -\frac{M_{\rm p}}{M_{\rm trans}} \right),
%\label{mdot_app}
%\end{equation}
%where $C$ is a constant.  Integrating eq.[\ref{mdot_app}], we obtain
%\begin{equation}
%\dot{M}_{\rm p}
% \simeq \frac{M_{\rm trans}}{t},
% \label{mdot_analytic_1}
%\end{equation}
%when $t$ and $M_{\rm p}$ are large.

When $2r_{\rm H} \leq x_{\rm m}$, $\dot{M}_p$ depends on $M_p$ in a more
complicated manner.  As seen in equation (\ref{Sigma_R}), $\Sigma_R$ is
not a simple exponential function of $M_{\rm p}$ unlike $\Sigma_{\rm
vis}$.  Defining a function $f(M_p)$ as
\begin{equation}
f(M_{\rm p}) \equiv
 -\frac{1}{2}
  \left( \frac{2r_{\rm H}}{h} - \frac{5}{4}\frac{x_{\rm m}}{h}\right)^2
 +\frac{1}{32}
  \kakkoi{x_{\rm m}}{h}{2}
 -\kakkoi{x_{\rm m}}{\ell}{-3},
\end{equation}
one can write $\Sigma_{\rm acc}$ in the form
\begin{equation}
\Sigma_{\rm acc} =
\Sigma_{\rm R}(2r_{\rm H})
 = \Sigma_\infty \exp(f(M_{\rm p})).
\end{equation}
Since $M_p$ does not change so much in the post-gap phase, one only has
to integrate $\dot{M}_p$ in a limited range of $M_p$.
Referring to $\Delta M_p$ as an increment from a mass, $M_{\rm p,0}$
(i.e., $M_{\rm p} = M_{\rm p,0} + \Delta M_{\rm p}$), one expands $e^f$
in terms of the small quantity $\Delta M_p$ in such a way
\begin{eqnarray}
\Sigma_{\rm acc} =
\Sigma_{\rm R}(M_{\rm p,0} + \Delta M_{\rm p})
&=& \Sigma_\infty
    \exp\left( f(M_{\rm p,0}) + \frac{df}{dM_{\rm p}} \Delta M_{\rm p}
        \right) \nonumber \\
&=& \Sigma_{\rm R}(M_{\rm p,0})
    \exp\left( -\frac{\Delta M_{\rm p}}{(-df/dM_{\rm p})^{-1}}
        \right),
\end{eqnarray}
where
\begin{equation}
\frac{df}{dM_{\rm p}}
 = \frac{1}{M_{\rm p}}
   \left(
   -\frac{1}{3}
    \kakkoi{2r_{\rm H}}{h}{2}
   +\frac{11}{12^{4/5}}
    \kakkoi{2r_{\rm H}}{h}{}
    \kakkoi{\ell}{h}{3/5}
   -\frac{8}{12^{3/5}}
    \kakkoi{\ell}{h}{6/5}
   \right).
 \label{dfdMp}
\end{equation}
Thus one obtains 
\begin{equation}
\dot{M}_{\rm p}
 \simeq \frac{(-df/dM_{\rm p})^{-1}}{t},
 \label{mdot_analytic_2}
\end{equation}
for large $t$ and $M_{\rm p}$.




\clearpage

%% the file, uncomment the \plotone, \plottwo, and
%% If you need a layout that cannot be achieved with \plotone or
%% \plottwo, you can invoke the graphicx package directly with the
%% \includegraphics command or use \plotfiddle. For more information,
%% please see the tutorial on "Using Electronic Art with AASTeX" in the
%% documentation section at the AASTeX Web site,
%% http://www.journals.uchicago.edu/AAS/AASTeX.
%%
%% The examples below also include sample markup for submission of
%% supplemental electronic materials. As always, be sure to check
%% the instructions to authors for the journal you are submitting to
%% for specific submissions guidelines as they vary from

%% This example uses \plotone to include an EPS file scaled to
%% 80% of its natural size with \epsscale. Its caption
%% has been written to indicate that additional figure parts will be
%% available in the electronic journal.

\clearpage

\clearpage

\begin{deluxetable}{lll}
\tabletypesize{\scriptsize}
%\rotate
\tablecaption{List of notations\label{table-1}}
\tablewidth{0pt}
\tablehead{
\colhead{Variables} & \colhead{Meaning} & \colhead{Definition} }
\startdata
$M_{\rm p}$ & Planetary mass & \\
$M_\ast$ & Mass of the central star& \\
$M_{\rm disk}$ & Disk mass at the initial condition & $\int_0^\infty
 2\pi r \Sigma_{\rm ss}(r,0) dr$\\
$M_{\rm final}$ & Final mass of a planet & Eq.[\ref{M_final}]\\
$M_{\rm final,gap}$ & Final mass of a planet in the gap-limiting region & Eq.[\ref{M_final_gap}]\\
$M_{\rm final,diff}$ & Final mass of a planet in the diffusion-limiting
 region & Eqs.[\ref{M_final_diff_1}] and [\ref{M_final_diff_2}]\\
$M_{\rm trans}$ & Planetary mass when $\Sigma_{\rm acc}$ is
 $(1/e)\Sigma_\infty$ in the case of $2r_{\rm H} \geq x_{\rm m}$ & Eq.[\ref{M_trans}]\\
$M_{\rm p,init}$ & Initial planetary mass & \\
$M_{\rm p,crit}$ & Maximum planetary mass that does not violate
 a geometrical limit for eq.[\ref{Adot}] & Eq.[\ref{M_p_crit}] \\
$M_{\rm p,local}$ & Final mass if $\dot{M}_{\rm p}$ is assumed to be
 $\dot{M}_{\rm p,local}$ & Eq.[\ref{M_p_local-disk}] \\
$M_{\rm p,disk}$ & Final mass if $\dot{M}_{\rm p}$ is assumed to be
 $\dot{M}_{\rm disk}$ & Eq.[\ref{M_p_local-disk}] \\
$M_{\rm p,vis}$ & Planetary mass when viscous condition for gap opening
 is fulfilled & Eq.[\ref{M_p_vis}] \\
$\dot{M}_{\rm p}$ & Accretion rate onto a planet & Eqs.[\ref{mdot}] or
 [\ref{Mdot_p_min}] \\
$\dot{M}_{\rm p,local}$ & Accretion rate onto a planet using
 eqs.[\ref{Sigma_general}],[\ref{mdot}], and [\ref{Sigma_infty_ss}]
 & see \S \ref{sec_model} \\
$\dot{M}_{\rm disk}$ & Radial mass flux of the disk due to viscous evolution & Eq.[\ref{flux_self_similar}] \\
$\Sigma_\infty$ & Unperturbed surface density at planet's orbit & Eqs.[\ref{Sigma_vis}],[\ref{Sigma_R}],[\ref{Sigma_infty_init}],[\ref{Sigma_infty_ss}]\\
$\Sigma_{\rm acc}$ & Surface density at the accretion band ($x=2r_{\rm H}$) & Eq.[\ref{mdot}]\\
$\Sigma_{\rm vis}$ & Surface density determined by the balance between
 viscous and gravitational torques & Eq.[\ref{Sigma_vis}]\\
$\Sigma_{\rm R}$ & Surface density determined by marginally stable state
 of the Rayleigh condition & Eq.[\ref{Sigma_R}]\\
$\Sigma_{\rm ss}$ & Surface density of self-similar solution for viscous
 evolving disk & Eq.[\ref{Sigma_self_similar}]\\
$\Sigma_{\infty, \rm init}$ & Initial unperturbed surface density at planet's orbit & Eq.[\ref{Sigma_infty_init}]\\
$r_{\rm p}$ & Semi-major axis of a planet& \\
$r_{\rm b}$ & Boundary position between the gap-limiting region and
 the diffusion-limiting region & see \S \ref{sec_classification}\\
$r_{\rm e}$ & Boundary position between the diffusion-limiting region
 and the no-growth region & see \S \ref{sec_classification}\\
$\ell$ & Position where $\Sigma_{\rm vis}$ is $(1/e)\Sigma_\infty$ &
 Eqs.[\ref{Sigma_vis}] or [\ref{ell}]\\
$x_{\rm m}$ & Position where the Rayleigh condition is marginally
 fulfilled for $\Sigma_{\rm vis}$ & Eq.[\ref{x_m}]\\
$\tau_{\rm div}$ & Time when the accretion rate is the maximum & Eq.[\ref{tau_div}]\\
$\tau_{\rm dep}$ & Exponential depletion timescale of a disk & Eq.[\ref{Sigma_infty_init}]\\
$\tau_{\rm vis}$ & Viscous evolution timescale of a disk & Eq.[\ref{tau_vis}] \\
$\tau_{\rm lifetime}$ & Effective disk lifetime (i.e., shorter one of
 $\tau_{\rm dep}$ and $\tau_{\rm vis}$)& Eq.[\ref{tau_lifetime}]\\
$\zeta$ & Normalized parameter (Surface density times disk depletion
time) & Eq.[\ref{zeta}]\\
\enddata
%% Text for table notes should follow after the \enddata but before
%% the \end{deluxetable}. Make sure there is at least one \tablenotemark
%% in the table for each \tablenotetext.
%\tablecomments{}
%\tablenotetext{b}{Another sample footnote for table~\ref{tbl-1}}
\end{deluxetable}

\end{document}